\documentclass[pre,twocolumn,nofootinbib]{revtex4-1}
\usepackage{amsmath,amssymb}    
\usepackage{graphicx}
\usepackage{bm}
\usepackage{mathrsfs}
\usepackage{comment}
\makeatletter

\begin{document}

\title{Scaling laws for frictional granular materials confined by constant pressure under oscillatory shear}
\author{Daisuke Ishima}
\email[]{daisuke.ishima@yukawa.kyoto-u.ac.jp}
\author{Hisao Hayakawa}
\affiliation{Yukawa Institute for Theoretical Physics, Kyoto University, Kyoto 606-8502, Japan}
\date{\today}
\begin{abstract}
Herein, we numerically study the rheology of a two-dimensional frictional granular system confined by constant pressure under oscillatory shear.
Several scaling laws for the storage and loss moduli against the scaled strain amplitude have been found. 
The scaling laws in plastic regime for large strain amplitude can be understood by the angular distributions of the contact force.
The scaling exponents are estimated by considering the physical mechanism.
\end{abstract}
\maketitle

\section{Introduction\label{intro}}
Amorphous materials consisting of repulsive and dissipative grains such as granular materials~\cite{Gra}, colloidal suspensions \cite{Col}, bubbles \cite{For}, and emulsions \cite{Zhang05} exhibit characteristic viscoelastic properties.
Such materials behave as solids if the grains are densely packed, 
while they behave as liquids if the grains are not sufficiently dense.
There is a jamming transition at a critical density where the viscosity diverges and the rigidity emerges above the critical density at zero load limit.
Since the pioneer work by Liu and Nagel \cite{LNjam}, 
studies on the jamming transition have attracted considerable attention among physicists.
The jamming transition for frictionless grains is known as a mixed transition in which the coordination number changes discontinuously
while the pressure and the shear stress change continuously at the jamming point \cite{Djam,Mjam1,Tjam,Ojam,Mjam3,Mjam}. 
Such a continuous change of the stress near the jamming point satisfies a number of scaling laws \cite{Esca1,Osca, Hsca, Wsca2, Esca2, OHsca}.

It is remarkable that the introduction of mutual frictions between the grains drastically changes the behavior of the jamming transition including the existence of a discontinuous shear thickening (DST)~ \cite{SHsca,Ssca2,dst1,Sisca, OHdst1, SMdst, dst2, OHdst2, Kdst}.
Moreover, the mutual friction cannot be ignored for rigid grains to be consistent with Newton's equation of motion.
We have recognized that shear jammed states,  commonly observed in frictional systems, differ from the conventional isotropic jammed states \cite{Zsj,Exsj1,Exsj2,Exsj3,Nusj1,Nusj2,OHsj,Pradipto}.
Several reserchers are interested in the mutual relationship between shear jamming and DST~\cite{OHsj,Pradipto,Fall15,Peters16}.
We realize that some shear jamming has originated from a memory effect of mechanical training~\cite{OHsj,Pradipto,Kumar,Jin17,Urbani,Jin18}.

If an oscillatory shear is applied to such a material, the rigidity and viscosity can be measured simultaneously \cite{LAOS1,LAOS2,HyunRev,LAOS3}.
Recently, scaling laws on the rigidity and viscosity for frictionless grains under oscillatory shears have been studied \cite{Tosc,Otsuki14,BBosc,BSosc,BTosc}.
In particular, it is interesting that two distinct regions exist, softening and yielding, when considering large strain amplitude~\cite{BSosc}.  
Previous studies have discussed scaling laws for frictional grains under simple shear~\cite{OHdst1} and oscillatory shear~\cite{OHdst2}; 
however, thus far, most of these studies are based on situations under constant volume conditions.

Dilatancy is widely known as one of the typical characteristics of granular systems in which the density decreases as the shear increases under constant pressure conditions \cite{Rdil,Tdil,Sdil,MiDi04,Fdil,Bdil}.
Because the dilatancy cannot be observed in a constant volume system,
it is important to understand the rheological properties of granular systems confined by the constant pressure under oscillatory shear.

The purpose of this study is to determine the viscoelastic properties of an oscillatory sheared granular system. 
For this purpose, we numerically study a granular system confined by constant pressure under oscillatory shear to extract scaling laws for the storage and loss moduli.
The results suggest the absence of the distinction between softening and yielding in frictional systems, 
which contrasts with that of frictionless cases~\cite{BSosc}.

The contents of this paper are as follows.
In the next section, we explain the setup of our numerical simulation.
Section~\ref{sec3} illustrates the relevance of several scaling laws for the storage and loss moduli with a fixed friction constant for $\mu=1$, where $\mu$ is Coulomb's friction constant.
In the plastic regime  under large strain amplitude, 
the semi-quantitative behavior of the stress can be understood by the consideration of the angular distributions of contacts. 
In Sec. \ref{Apppsca} we demonstrate that the scaling laws observed for $\mu=1$ in Sec. \ref{sec3} are essentially independent of $\mu$.
In the last section, we discuss and summarize our results.
In the appendices, we explain the technical details and present some supplemental information which has not been described in the main text.

\section{simulated system\label{sec2}}

Our system is a two-dimensional one containing $N$ frictional granular disks.
To avoid crystallization, we adopt a tetradisperse system in which the numbers of grains for the diameters  $d_{0},\ 0.9d_{0},\ 0.8d_{0},$ and $0.7d_{0}$ are $N/4$~\cite{Crys}.
We assume that the density of each grain is identical and, thus, the mass is proportional to the square of the diameter of the grain.
The mass corresponding to the largest diameter $d_{0}$ is denoted by $m_{0}$ in this study.

\begin{figure}[htbp]
  \centering
    \includegraphics[clip,width=8.5cm]{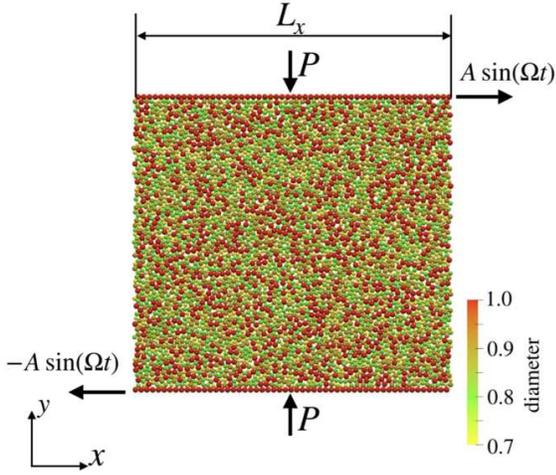}
    \caption{Image of simulated system, where $P$, $A$ and $\Omega$ are external pressure, strain amplitude, and angular frequency of external oscillation, respectively.}
    \label{fig:setup}
\end{figure}

We assume that gravity is negligible in our system corresponding to grains on a smooth horizontal plane.
We adopt the periodic boundary condition in the shear ($x-$)direction, 
while we apply the external pressure through the wall in the $y-$direction.
The grains are confined in a square box, i.e. $L_{0}:=L_y=L_x$ in the absence of any external pressure,
where $L_x$ and $L_y$ are the linear dimensions in the $x-$ and $y-$directions, respectively.
When the external pressure is applied, $L_y(t)$ is no longer a constant but depends on time.
The position of the center of mass of the wall at $y=\pm L_{y}(t)/2$ in the $x-$direction $x_{G}^{+}(t)$ obeys $x_{G}^{\pm}(t)= \pm A\sin(\Omega t)$,
where $A$ and $\Omega$ are the strain amplitude and the angular frequency, respectively (see Fig. \ref{fig:setup}).
Because the linear dimension of the system size in the $y-$direction is not constant, we introduce the effective strain amplitude
\begin{align}
\gamma_{0,\textrm{eff}}:=\frac{2A}{L_{0}} .
\end{align}

To describe the motion of the walls under the external pressure $P$, we adopt the following equation:
\begin{align}
m_{w}\frac{dv_{w,y}^{\pm}}{dt}=\pm(P_{w}^{{\pm}} - P)L_{x}-\xi_{\textrm{d}}  v_{w,y}^{\pm}\label{eq:walls},
\end{align}
where $\pm$, $m_w$, $v_{w,y}^{\pm}$, $P_w^{\pm}$, and $\zeta_{\textrm{d}}$ are the superscripts for the wall at $y=\pm L_{y}(t)/2$, the mass of the wall satisfying $m_w=N_w m_0$, $y-$component velocity of the wall at $y=\pm L_y(t)/2$, the inner pressure acting on the walls from the grains, and the damping constant, respectively.

The translational and rotational equations of the motion of the $i-$th grain whose mass, position, and moment of inertia are, respectively, $m_i$, $\bm{r}_i$, and $ I_i=m_i d_i^2/8$, respectively, with the diameter $d_i$ are given by
\begin{align}
m_i \frac{d^{2}\bm{r}_i}{dt^{2}}=\sum_{j\neq i}\bm{f}_{ij},\\
I_i\frac{d\omega_i}{dt}=\sum_{j\neq i}T_{ij},\label{eq:eqI}
\end{align}
where we have introduced the contact force $\bm{f}_{ij}$ and the torque $T_{ij}$ from the $j-$th grain acting on the $i-$th grain.
Note that $\omega_i$ is the $z-$component of the angular velocity of the $i-$th grain.
The torque $T_{ij}$ introduced in Eq. (\ref{eq:eqI}) satisfies
\begin{align}
T_{ij}=-\frac{d_i}{2}\bm{f}_{ij}\cdot\bm{t}_{ij}\label{eq:mom}
\end{align}
with the tangential unit vector $\bm{t}_{ij}$ at the contact point between the $i-$th and $j-$th grains.
The contact force of each grain is described by the Cundall-Stack model \cite{dem1, dem2}.
Considering the contact force $ \bm{f}_{ij}$ between the $i-$th and $j-$th grains,
we inntroduce $ d_{ij} = (d_i + d_j) / 2$ and $\bm {r}_{ij} = \bm{r}_i-\bm{r}_j$ with $r_{ij}:=|\bm{r}_{ij}|$, the contact force $\bm{f}_{ij}$ is expressed as
\begin{align}
\bm{f}_{ij}=\left(\bm{f}_{ij,n}+\bm{f}_{ij,t}\right)\Theta(d_{ij}-r_{ij}),\label{eq:fnt}
\end{align}
where we have introduced the normal contact force $\bm{f}_{ij,n}$, the tangential contact force $\bm{f}_{ij,t}$ and Heaviside's step function $\Theta(x)$ satisfying $\Theta(x)=1$ for $x\ge 0$ and $\Theta(x)=0$ otherwise. The normal force consists of the elastic repulsive force represented by a linear spring (the spring constant $k_n$) and the dissipative force represented by a dashpot (the damping constant $\xi_n$) as
\begin{align}
\bm{f}_{ij,n}&=k_n\delta_{ij,n}\bm{n}_{ij}-\xi_n\bm{v}_{ij,n},\label{eq:fn}
\end{align}
where we have introduced $\bm{n}_{ij}=\bm{r}_{ij}/r_{ij}$, $\bm{v}_{ij,n}=(\bm{v}_{ij}\cdot\bm{n}_{ij})\bm{n}_{ij}$ with $\bm{v}_{ij}=d\bm{r}_{ij}/dt$ and $\delta_{ij,n}=d_{ij}-r_{ij}$.
On the other hand, the tangential force $\bm{f}_{ij,t}$ contains both a sticking state and a slip state,
where the switching from the stick to the slip takes place if the magnitude of the tangential force $f_{ij,t}=|\bm{f}_{ij,t}|$ exceeds the critical condition as
\begin{align}
  \bm{f}_{ij,t} &= \begin{cases}
    k_t\delta_{ij,t}\bm{t}_{ij}-\xi_t\bm{c}_{ij} & ( f_{ij,t}<\mu f_{ij,n} ) ,\\
    \mu f_{ij,n}\bm{t}_{ij} & (\textrm{otherwise})\label{eq:fri},
  \end{cases}
\end{align}
where $\delta_{ij,t}$ is the tangential displacement during a contact time.
We have also introduced Coulomb's friction coefficient $\mu$, the tangential spring constant $k_t$, the tangential damping constant $\xi_t$, $f_{ij,n}=|\bm{f}_{ij,n}|$ and the relative tangential velocity at the contact point $\bm{c}_{ij}:=\bm{v}_{ij}-\bm{v}_{ij,n}+\bm{t}_{ij}(d_i\omega_i+d_j\omega_j)/2$.

For our simulation, we adopt $k_{t}=k_{n}/2$, $\xi_{n}=(m_{0}k_{n})^{-1/2}$, $\xi_{t}=\xi_{n}$, and $\xi_{\textrm{d}}=\xi_{n}$ which follow Refs. \cite{OHsj,Cruz}.
The control parameters of our simulation are $\hat{P}:=P/k_n$, $\gamma_{0,\rm{eff}}$ and $\Omega\sqrt{m_0/k_n}$.
We cover the range of $\hat{P}$ from $2.0\times10^{-5}$ to $6.0\times10^{-3}$.
In most of our simulations we adopt the number of grains $ N=4,000 $, 
and the frequency $ \Omega/(2 \pi)= 1.0\times10^{- 4} \sqrt{k_{n}/m_{0}} $.
We have confirmed that the rigidity and the viscosity are independent of $N$ in the range of $4,000\leq N\leq20,000$.
We have checked the dependence of $\Omega$ in the range of $4.0\times10^{-5}\sqrt{k_{n}/m_{0}}\leq\Omega/(2 \pi)\leq1.0\times10^{-3}\sqrt{k_{n}/m_{0}}$ (see Appendix \ref{Appome}) and confirmed that the storage modulus is independent of $\Omega$ and the loss modulus depends on $\Omega$ in the small strain regime but does not depend on $\Omega$ for large strain.

To make the initial configuration isotropic,
we randomly place all grains whose diameters are $ 0.6d_{0},\ 0.5d_{0},\ 0.4d_{0}, \textrm{and}\ 0.3d_{0}$ corresponding to tetradispersity of grains confined in a square box with the initial density $\phi_{\textrm{ini}}=0.24$ without any overlap and shear.
After we increase the diameter of each grain by $ 0.02 d_ {0} $, we wait for the system to relax to a steady state in which the potential energy $V_{n}:=k_n\sum_{i\neq j}\delta_{ij,n}^{2}/2$ for the normal compression satisfies $ V_ {n} /(N k_ {n} d_ {0} ^ {2}) <2.0 \times10 ^{-7} $.
Subsequently, we increase the diameter of each grain by a further $ 0.02 d_ {0} $ again.
After repeating this process, we reach a desired state in which the area fraction is $\phi=0.82$.
We compress both walls by pressure $P$ to compactify the system above the jamming density to achieve an isotropic structure as the initial condition.
After $V_{n}$ converges to the steady value, 
we apply oscillatory shear.
We adopt the symplectic Euler method with the time step $\Delta t=0.05\sqrt{m_{0}/k_{n}}$.

We average the data over $N_T$ cycles after abandoning the data in the initial $N_{\textrm{ini}}$ cycles.
Because the obtained results depend on $N_{\textrm{ini}}$ for $N_{\textrm{ini}}\le 5$, we adopt $N_{\textrm{ini}}=10$ and $N_{T}=10$.
Furthemore, we have confirmed that the storage and loss moduli are independent of $N_{\rm ini}\leq 190$ except for the low pressure cases and the critical region of the yielding transition.
We set $t = 0$ at the instant when $N_{\textrm{ini}}$ oscillatory shear ends.

\begin{figure}[htbp]
  \centering
    \includegraphics[clip,width=8.5cm]{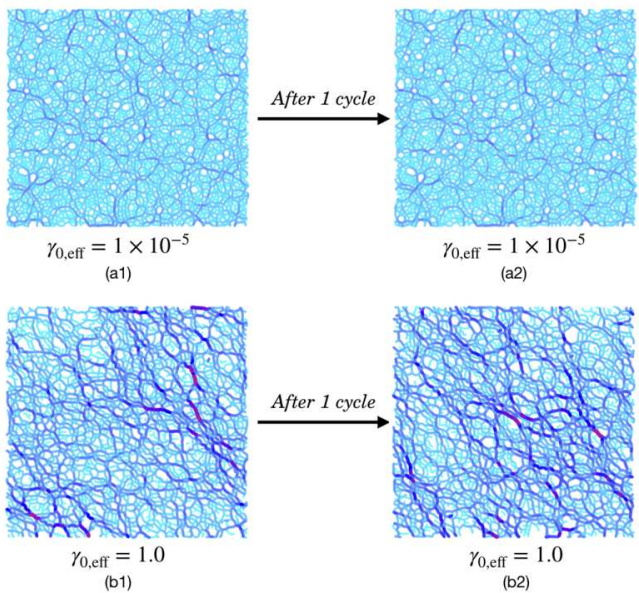}
    \caption{Time evolution of force chains during one cycle at $\hat{P}=2.0\times10^{-3},\ \mu=1.0,\ \textrm{and}\ N=4,000$ for $\gamma_{0,\textrm{eff}}=1.0\times 10^{-5}$ (top) and $\gamma_{0,\textrm{eff}}=1.0$ (bottom).
    Width of each line is proportional to the absolute value of the contact force between grains. }
    \label{fig:fcs}
\end{figure}

Figure~\ref{fig:fcs} shows the time evolution of the force chains during one cycle in which the force chains are unchanged if the strain amplitude is small;
however, they are changed if the strain amplitude is large.
The former regime, in which the granular materials behave as an elastic solid, is elastic, 
while the latter one, in which the granular materials behave as a plastic material with configuration changes of the grains, is plastic.
The yielding transition is the transition between the two states.

Because the contact stress is dominant for the stress in dense granular systems \cite{Cruz,rotTen}, 
we obtain the symmetric contact stress $\sigma_{\alpha\beta}^{\textrm{sym}}$ as
\begin{align}
\sigma_{\alpha\beta}^{\textrm{sym}}:=-\frac{1}{2L_{x} L_{y}}\sum_{i}\sum_{j>i}(x_{ij,\alpha}f_{ij,\beta}+x_{ij,\beta}f_{ij,\alpha})
\label{eq:sig}.
\end{align}
The asymmetric part in the stress tensor is associated with the coupled stress in frictional systems~ \cite{Mitarai02,Gol}.
Nevertheless, we have confirmed that the asymmetricity of the stress tensor is so small that we can ignore the asymmetric stress tensor or the coupled stress (see Appendix \ref{Appasym}).
We note that the effect of the shear band near the boundaries is not negligible for low pressure and large strain amplitude.
In this case, the measured stress from Eq.~\eqref{eq:sig} might be different from the local stress.

In systems under oscillatory shear, the storage modulus $ G'$ and the loss modulus $ G''$ represent the rigidity and dynamic viscosity multiplied by $\Omega$, respectively.
Here, the shear stress $\sigma(t):=\sigma_{xy}^{\textrm{sym}}$ can be decomposed into the elastic and viscous parts as
\begin{align}
\sigma(t)&=\sigma^{\textrm{(ela)}}(t)+\sigma^{\textrm{(vis)}}(t),
\end{align}
where $\sigma^{\textrm{(ela)}}(t)$ and $\sigma^{\textrm{(vis)}}(t)$ are expressed as $\sigma^{\textrm{(ela)}}(t)=G'\gamma(t)$ and 
$\sigma^{\textrm{(vis)}}(t)=\eta_d \dot{\gamma}(t)$
with the strain $\gamma(t)$ and the strain rate $\dot\gamma(t)$, respectively.
It is known that 
the dynamic viscosity can be expressed as the loss modulus $G''$ as $\eta_d:=G''/\Omega$. 
It should be noted that their definitions are not unique in the nonlinear response regime \cite{LAOS1, LAOS2,HyunRev, LAOS3}.
In this study, we adopt the following expressions \cite{LAOS3}:
\begin{align}\label{G'}
G'(\gamma_{0,\textrm{eff}}, P)&:=\lim_{\gamma(t)\rightarrow\gamma_{0,\textrm{eff}}}\frac{\tilde\sigma}{\gamma_{0,\textrm{eff}}},\\
G''(\gamma_{0,\textrm{eff}}, P)&:=\lim_{\gamma(t)\rightarrow0 (\tilde\sigma\geq0)}\frac{\tilde\sigma}{\gamma_{0,\textrm{eff}}},
\label{G''}
\end{align}
where we have introduced $\gamma(t):=\gamma_{0,\textrm{eff}}\sin(\Omega t)$ 
and $\tilde{\sigma}$ is defined as
\begin{align}
\tilde\sigma(t):=\sigma_{xy}^{\textrm{sym}}(t)-\overline{\sigma_{xy}^{\textrm{sym}}}\label{eq:fluc}.
\end{align}
Here $\overline{\sigma_{xy}^{\textrm{sym}}}:=\int_{0}^{\tau_p} dt\sigma_{xy}^{\textrm{sym}}(t)/\tau_p$ is the time average over the oscillatory period $\tau_p:=2\pi/\Omega$.
\footnote{
We have examined another protocol in which the
oscillatory shear is given by $\gamma(t)=\gamma_{0,\textrm{eff}}(1-\cos(\Omega t))$, and found that
the results of $G'$ and $G''$ are unchanged for this oscillation except for the region near the bending point of $G'$.
The choice of $\gamma(t)$ is important to discuss the fragile phase~\cite{OHsj,Pradipto}; 
however, we are not interested in the fragile phase. 
}

\section{Scaling laws for storage and loss moduli for $\mu=1$~\label{sec3}}

In this section we illustrate the existence of several scaling laws for the storage and loss moduli at $\mu=1.0$.
The results are basically independent of $\mu$ for $0.01\le \mu\le 1$ as will be shown in Sec. \ref{Apppsca}.
We also demonstrate that the stress-strain curve observed in the plastic regime can be reproduced by a phenomenology.

\subsection{Scaling law for the storage modulus}

\begin{figure}[htbp]
  \centering
    \includegraphics[clip,width=8.5cm]{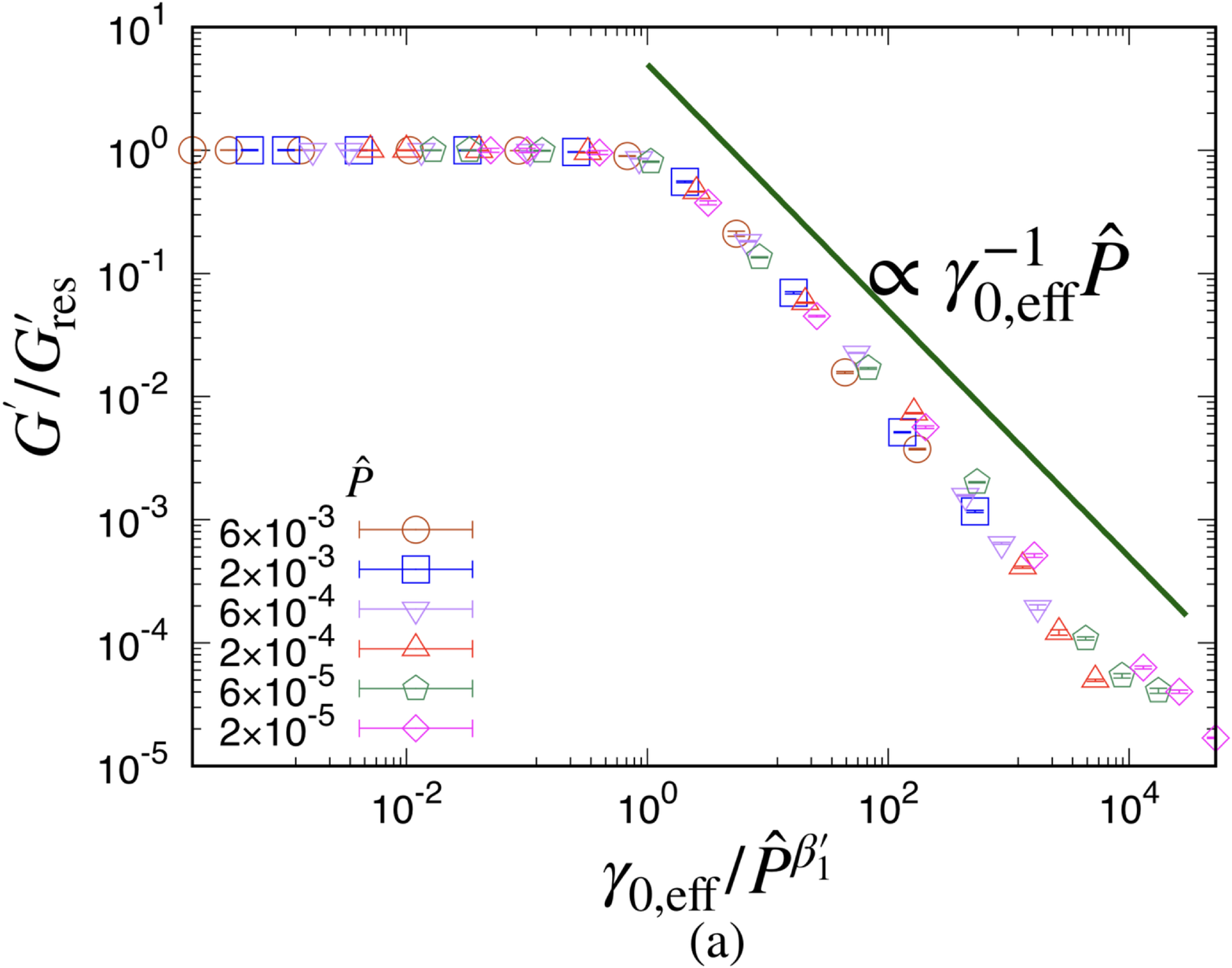}
    \includegraphics[clip,width=8.5cm]{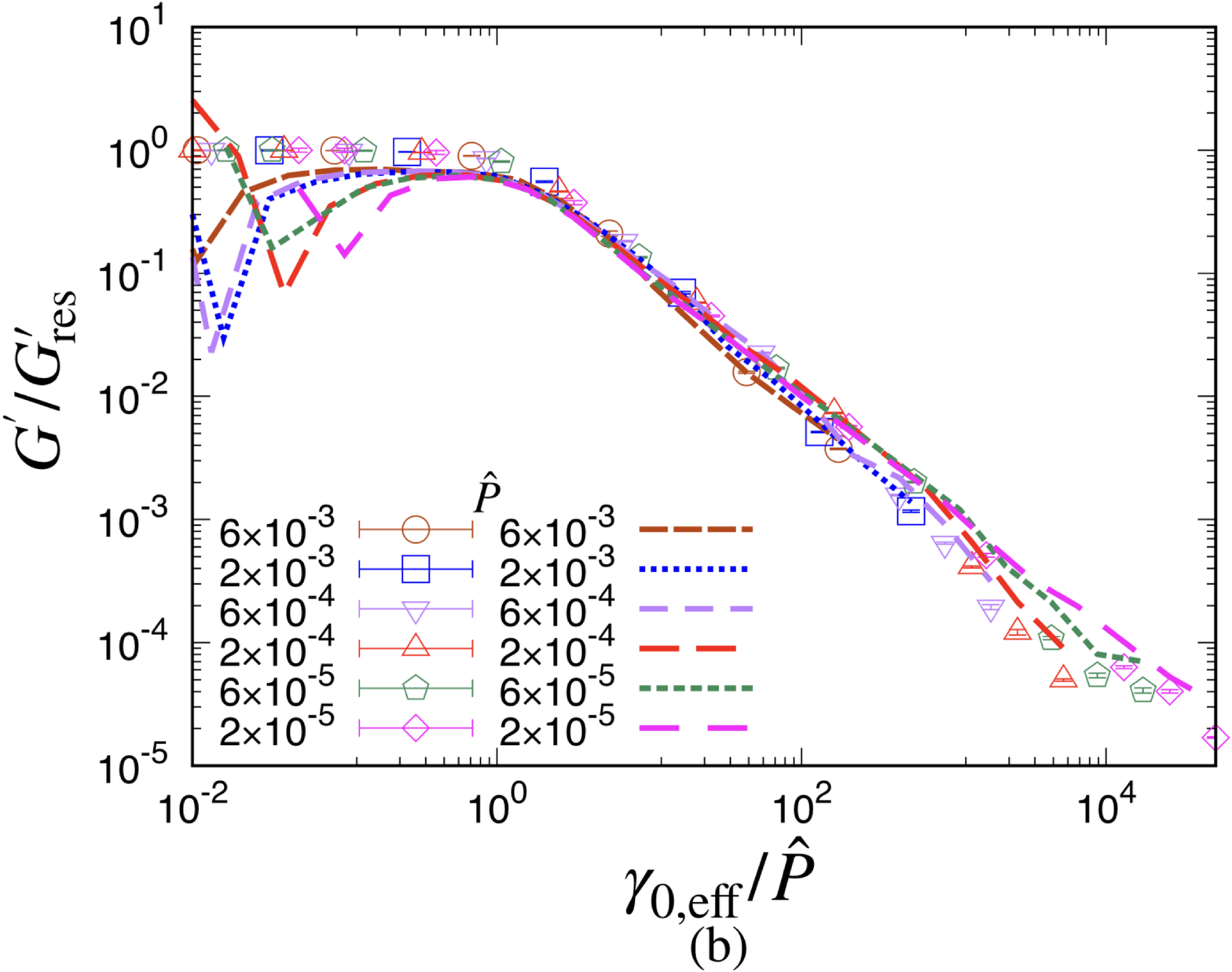}
    \caption{
   (a)  Scaling plots of storage modulus $G'/G'_{{\textrm{res}}}$ for various dimensionless pressures $\hat P$ against the scaled $\gamma_{0,\textrm{eff}}$ at $\mu=1.0$, where $G_{\textrm{res}}'$ is residual storage modulus.
     (b) Comparison of phenomenological storage modulus (lines) with numerical ones (data) for $\gamma_{0,\textrm{eff}}/\hat P\ge 0.01$.
    }
    \label{fig:sGs}
\end{figure}

From our simulation,
we find the following scaling form for the storage modulus
(see Fig. \ref{fig:sGs}):
\begin{align}
G'&=G'_{\textrm{res}}( \hat P)\mathcal{G'}\left( \frac{\gamma_{0,\textrm{eff}}}{\hat P^{\beta_{1}'}} \right),\label{eq:sgs1}\\
G_{\textrm{res}}'(\hat P):&=\lim_{\gamma_{0,\textrm{eff}}\to 0}G'(\gamma_{0,\textrm{eff}}, \hat P), \label{eq:sgs2}\\
\lim_{x\to 0}\mathcal{G'}(x)&=1,\ \lim_{x\rightarrow\infty}\mathcal{G'}(x)\sim x^{-1} .
\label{eq:sgs3}
\end{align}
The scaling exponent $\beta_1'$ which is a special case of the $\mu-$dependent exponent $\beta_\mu'$ is evaluated as
\begin{align}
\beta_{1}'=1.00\pm0.03. \label{eq:sgs4}
\end{align}
Note that $\beta_\mu'$ is basically independent of $\mu$ (see Sec. \ref{Apppsca}).

\begin{figure}[htbp]
  \centering
    \includegraphics[clip,width=8.5cm]{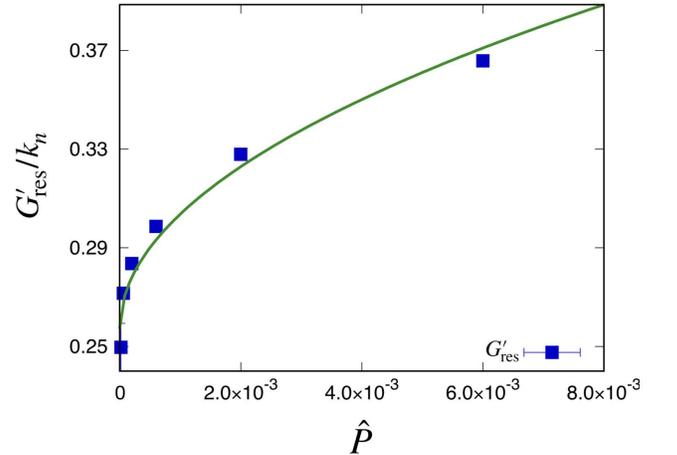}
    \caption{Plot of $G_{\textrm{res}}'$ defined by Eq. \eqref{eq:greseq} against $\hat{P}$ for $\mu=1.0$.
}
    \label{fig:Gres}
\end{figure}

Further, the residual storage modulus $G'_{\textrm{res}}(\hat P)$ in the low strain limit depends on the pressure as
\begin{align}
G'_{\textrm{res}}/k_{n}=
G'_{1}+a \hat P^{1/2} \label{eq:greseq} ,
\end{align}
where $G'_{1}=0.26\pm0.01$ and $a=1.49\pm0.12$ (see Fig.~\ref{fig:Gres}).
This behavior can be interpreted as follows.
It is widely known that the excess coordination number $\Delta z:=z-z_{\textrm{iso}}$ is proportional to $\sqrt{\hat{P}}$, i. e.  $\Delta z \propto \hat{P}^{1/2}$, where $z$ and $z_{\textrm{iso}}$ are the coordination and the isostatic coordination numbers, respectively~\cite{Ojam,Mjam,Djam,Hrev}. 
Because the shear stress is expected to be proportional to the excess coordination number in the elastic regime, we obtain $G'_{\textrm{res}}-G'_{\mu}\propto\hat P^{1/2}$, 
where $G_\mu'$ is a constant, irrespective of the pressure.

We find that $G'$ is independent of $\gamma_{0,\textrm{eff}}$ for small $\gamma_{0,\textrm{eff}}$, while $G'$ is proportional to $\gamma_{0,\textrm{eff}}^{-1}$ for large $\gamma_{0,\textrm{eff}}$.
These asymptotic forms in Eqs. (\ref{eq:sgs1})-(\ref{eq:sgs4}) can be interpreted as follows.
If the strain is small, the material behaves as the Hookean regime in which $G'$ must be independent of $\gamma_{0,\textrm{eff}}$.
On the other hand, the stress ratio $\sigma_{xy}/P$ approaches a constant $\mu_{\textrm{macro}}$ for the large strain (plastic) regime because the macroscopic dynamical friction constant $\mu_{\rm macro}$ is well-defined in granular materials, which leads to $\beta_{\mu}'=1$.
This simple description explains the crossover from the Hookean regime for small strains to the plastic regime of $G'$ for large strains.

We also note that this argument is independent of $\mu$.
Therefore, we expect that the scaling law characterized by Eqs. \eqref{eq:sgs1}-\eqref{eq:sgs3} is held for arbitrary $\mu$.
We can also evaluate the bending point of $G'$, which is the turning point from the Hookean regime to the plastic regime, as $\hat{P}/\gamma_{0,{\rm eff}}\sim 1$ by setting the macroscopic friction constant $\mu_{\rm macro}$ to be unity.
Suprisingly, this crude estimation of the location of the bending point is reasonable.
Indeed, the bending point appears to be located in the range $1< \gamma_{0,{\rm eff}}/\hat{P}<2$ which provides a good estimation of the bending point.

\subsection{Scaling law for the loss modulus \label{scaling_loss}}

\begin{figure}[htbp]
  \centering
    \includegraphics[clip,width=8.5cm]{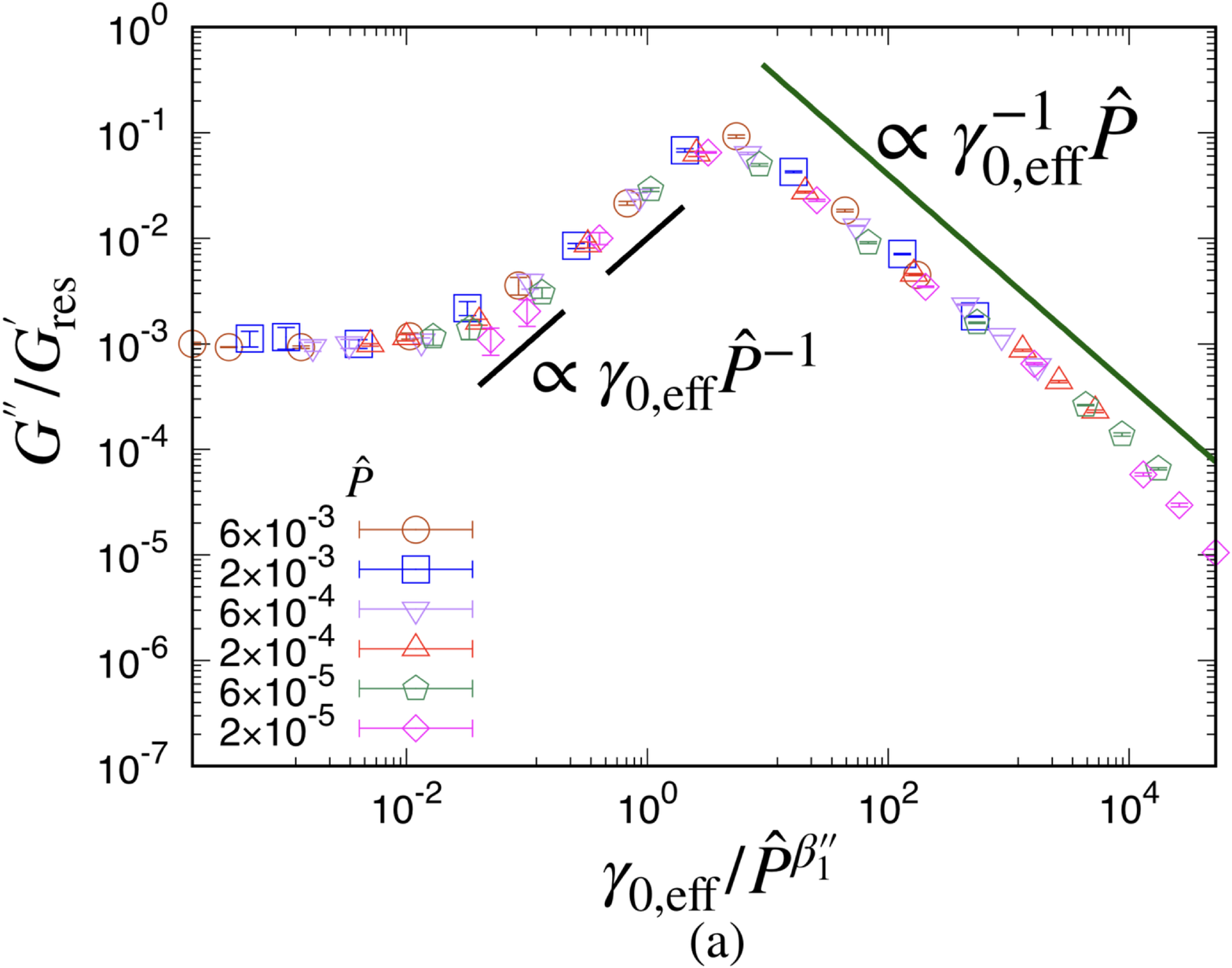}
    \includegraphics[clip,width=8.5cm]{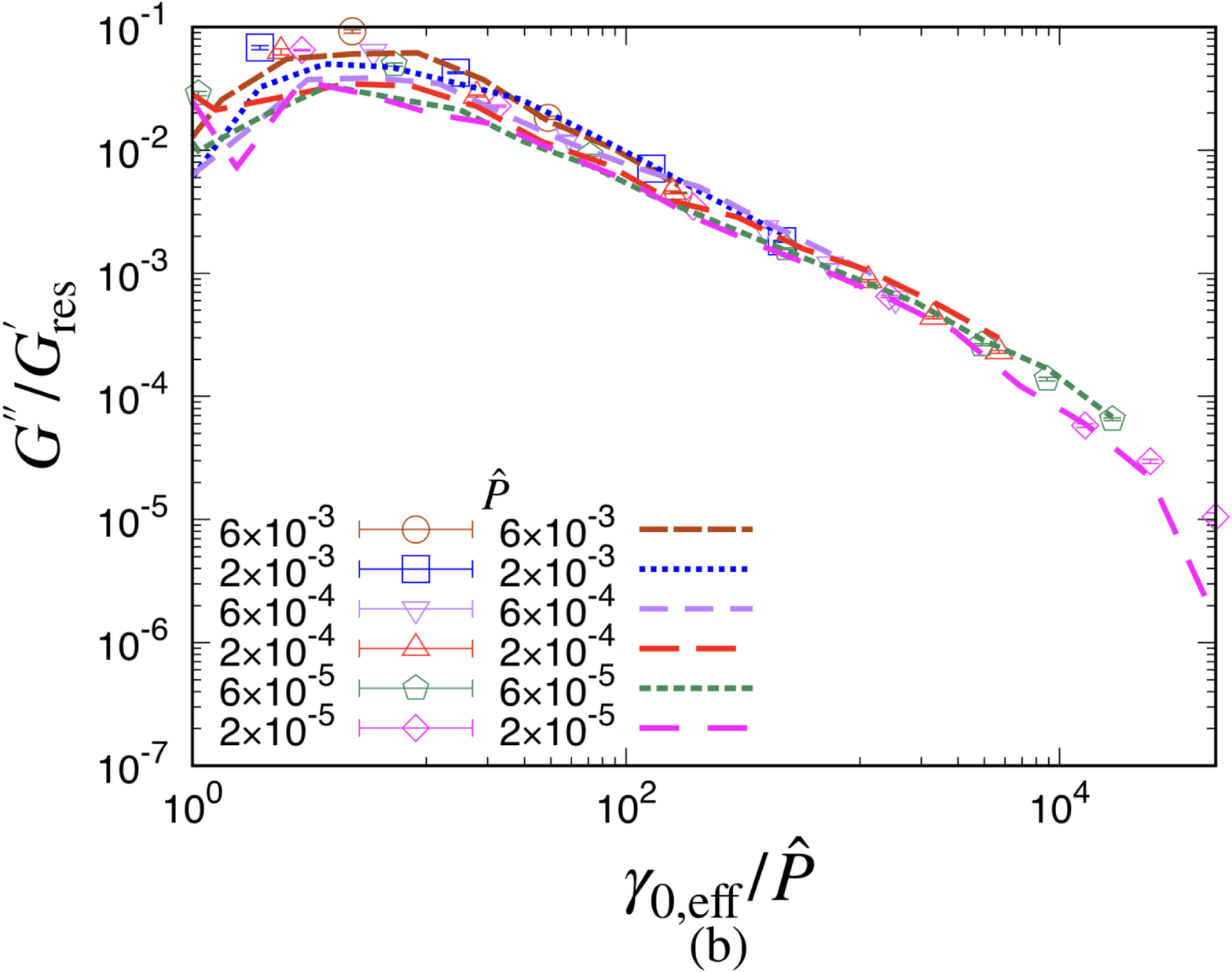}
    \caption{
(a) Scaling plots of loss modulus $G''$ for various $\hat{P}$ against the scaled strain amplitude at $\mu=1.0$, where $G''$ is proportional to $\gamma_{0,\textrm{eff}}^{-1}$ for large $\gamma_{0,\textrm{eff}}$.
     (b) Comparison of phenomenological loss modulus (lines) with numerical one (data)  for $\gamma_{0,\textrm{eff}}/\hat P\ge 1$.
}
    \label{fig:sGl}
\end{figure}

The loss modulus also satisfies the scaling form (see Fig.~\ref{fig:sGl}):
\begin{align}
G''&=
G'_{\textrm{res}}(\hat P)\mathcal{G''}\left( \frac{\gamma_{0,\textrm{eff}}}{\hat P^{\beta''_{1}}} \right), \label{eq:sgl1}\\
\lim_{x\to 0}\mathcal{G''}(x)&=\textrm{const.},\ \lim_{x\rightarrow\infty}\mathcal{G''}(x)\sim x^{-1},\label{eq:sgl3}
\end{align}
where the scaling exponents $\beta_1''$ which is a special cases of $\beta_\mu''$ and is evaluated as
\begin{align}
\beta_{1}''=1.05\pm0.01.\label{beta_1''}
\end{align}
We have confirmed that the  scaling exponent $\beta_{\mu}''$ is approaching independence of $\mu$ (see Sec. \ref{Apppsca}).

As with $G'$, the loss modulus in the plastic regime satisfies $G'' \sim P/\gamma_{0,\textrm{eff}}$ for large $\gamma_{0,\textrm{eff}}$ (i. e.  $\beta_\mu''=1$ )
because the stress ratio approaches a constant in this regime.
Figsures~\ref{fig:sGs} and \ref{fig:sGl} suggest that 
the bending point of $G'$ and the peak of $G''$ are located at $\gamma_{0,\textrm{eff}}/\hat P\simeq 1$ and $\gamma_{0,\textrm{eff}}/\hat P\simeq 4$, respectively. 
Although there is no reason for the location of the peak of $G''$ to be close to that of the bending point of $G'$, it is interesting that both points take place at the points $\gamma_{0,\textrm{eff}}/\hat P\sim O(1)$.
It is remarkable that $G''$ has three regimes:
(i) $G''$ is independent of $\gamma_{0, \textrm{eff}}$ for small strain amplitudes, 
(ii) $G''$ is linear for $\gamma_{0, \textrm{eff}}$ for the intermediate strain amplitudes, and
(iii) $G''\sim \hat{P}/\gamma_{0,\textrm{eff}}$ in the plastic regime for large $\gamma_{0, \textrm{eff}}$.
The existence of the linear regime (ii) suggests that the relaxation process for elastic vibrations plays an important role in this regime (see Appendix \ref{lossdim}).

To explain the crossover of $G''$ from regime (i) for small strain ($\gamma_{0,\textrm{eff}}/\hat P<x_{c}$) to regime (ii) for intermediate strain $x_c<\gamma_{0,\textrm{eff}}/\hat P<1$, where we have introduced the crossover point $x_c$.
Because the contact model of our simulation model is the Kelvin-Voight model,
$G''$ in regime (i) satisfies  (see Fig.~\ref{fig:Gslome} in Appendix \ref{Appome})
\begin{align}
G''/k_{n}\sim \hat\Omega \hat\xi_{n} \qquad (\gamma_{0,\textrm{eff}}/\hat P<x_{c}) ,
\label{eq:gls}
\end{align}
where $\hat\Omega=\Omega\sqrt{m_{0}/k_{n}}$ and $\hat\xi_{n}=\xi_{n}/\sqrt{m_{0}k_{n}}$.
On the other hand, $G''$ in the regime (ii) , as shown in Appendix \ref{lossdim}, is independent of $\Omega$ as
\begin{align}
G''/k_n\sim \frac{\gamma_{0,\textrm{eff}}}{\hat P}\qquad (x_{c}<\gamma_{0,\textrm{eff}}/\hat P<1).\label{eq:gll}
\end{align}
From the balance Eqs.~\eqref{eq:gls} and \eqref{eq:gll}, at $x_c$ we estimate $x_{c}\simeq \hat\Omega\hat\xi_{n}$, which is verified from our simulation (see Fig.~\ref{fig:xc} in Appendix \ref{Appome}).

\subsection{The role of the angular distribution functions of contact forces}

\begin{figure}[htbp]
  \centering
    \includegraphics[clip,width=8.5cm]{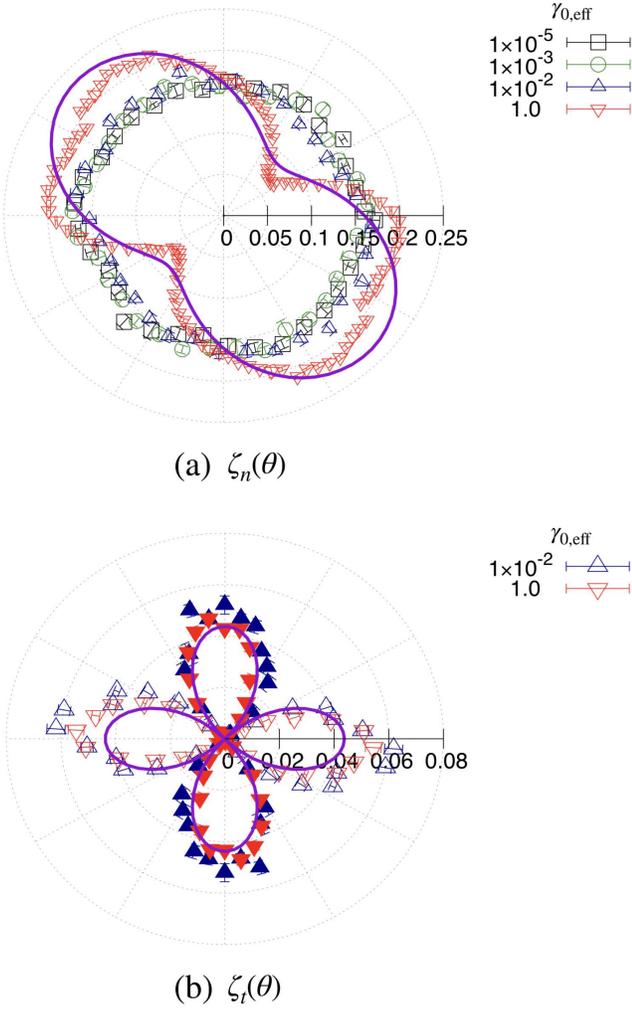}
    \caption{
 The angular distributions for (a) normal contact force density $ \zeta_ {N} (\theta) $ and (b) tangential contact force density $ \zeta_ {T} (\theta) $ at $ \Omega t = 2n\pi\ (n=1,\cdots, 9) $, $\mu=1.0$, and $\hat P=2.0\times10^{-3}$.
    Tangential force in clockwise direction is regarded as positive force, where positive forces are plotted as filled symbols and negative forces are plotted as open symbols.
    The purple lines in (a) $\zeta_{N}(\theta)$ and (b) $\zeta_{T}(\theta)$ are fitted by ${\rm const.} +b\sin(2\theta)$ and $c\cos(2\theta)$, respectively, at $\gamma_{0,\textrm{eff}}=1.0$.
     }
    \label{fig:afd}
\end{figure}

We have studied the behaviors of $G'$ and $G''$ as well as the stress-strain curve in the plastic regime quantitatively.
It is known that the approximate stress tensor can be expressed by the angular distributions of the normal contact force density $\zeta_{N}(\theta)$ and tangential contact force density $\zeta_{T}(\theta)$ in simple shear flows~\cite{Cruz,calten1,Kanatani,calten3}.
We try to apply this theory to grains under oscillatory shear.
Here $\zeta_{N}(\theta)$ and $\zeta_{T}(\theta)$ are expressed as $\zeta_{N}(\theta)=\rho(\theta)F_{N}(\theta)/\langle F_{N}\rangle$ and $\zeta_{T}(\theta)=\rho(\theta)F_{T}(\theta)/\langle F_{N}\rangle$ with the angular contact distribution $\rho(\theta)$,
the normal force $F_{N}(\theta)$, the tangential force $F_{T}(\theta)$ at the contact angle $\theta$, and the averaged contact force over the angle $\langle F_{\alpha}\rangle=\int_{0}^{2\pi} d\theta \rho(\theta)F_{\alpha}(\theta)$ for $\alpha=N$ or $T$.
There are the normalization conditions for these functions as $\int_0^{2\pi}d\theta \rho(\theta)=1$, $\int_{0}^{2\pi} d\theta \zeta_{N}(\theta)=1$ and $\int_{0}^{2\pi} d\theta \zeta_{T}(\theta)=0$, respectively.
Figure \ref{fig:afd} shows the plot of $\zeta_{N}(\theta)$ and $\zeta_{T}(\theta)$ in our system for various $\gamma_{0,\textrm{eff}}$ at $\hat P=2.0\times10^{-3}$.
In our system, the behavior of $\zeta_{N}(\theta)$ or $\zeta_{T}(\theta)$ for large strain is similar to that observed in simple shear \cite{Cruz,Kanatani}, while $\zeta_{N}(\theta)$ is isotropic for small strain.

In our system, even under oscillatory shear, the stress tensor $\sigma_{\alpha\beta}$ for large strain can be approximately expressed as
\begin{equation}
\sigma_{\alpha\beta} \approx
-2P\int_{0}^{2\pi} d\theta
\left[ \zeta_{N}(\theta)n_{\theta,\alpha} - \zeta_{T}(\theta)t_{\theta,\alpha} \right] {n}_{\theta,\beta}
\label{eq:theo}
\end{equation}
where we have used $P=-(\sigma_{xx}+\sigma_{yy})/2$.
Here, $(n_{\theta,\ x},\ n_{\theta,\ y})=(\cos\theta,\ \sin\theta)$
 and $(t_{\theta,\ x},\ t_{\theta,\ y})=(-\sin\theta,\ \cos\theta)$ are the normal and tangential unit vectors between contacting grains, respectively.
The phenomenological stress is only applicable to the plastic regime because the phenomenological stress is proportional to the pressure.

Using Eq. \eqref{eq:theo} our remaining task is to estimate $\zeta_N(\theta)$ and $\zeta_T(\theta)$ theoretically. 
The shear stress that originated from the normal contacts must be proportional to $\sin (2\theta)$ because the viscosity is proportional to $\int d\bm{r} \frac{xy}{r} V_n'(r) g_{\rm eq}(r)h(\bm{r})$ where $g_{\rm eq}(r)$ is the equilibrium radial distribution function and $h(\bm{r})$ is the shear contribution of the radial distribution function $g(\bm{r})=g_{\rm eq}(r)(1+\dot\gamma h(\bm{r}))$. 
Because of the orthogonality of the trigonometric functions, the angular distribution that contributes to the shear stress is proportional to $\sin (2\theta)$, while the equilibrium contribution should be independent of  $\theta$.
Similarly, the viscosity that originated from the tangential contacts contains $\cos^2\theta$ for the positive tangential contacts and $\sin^2\theta$ for the negative tangential contacts. 
As shown in Fig. \ref{fig:afd}, our simple description can recover the angular distributions of contacts with the introduction of fitting parameters $b$ and $c$.
Further, $b$ and $c$ depend on $\gamma_{0,\textrm{eff}}$ and $\hat P$ ( see Appendix \ref{fitting}).
We then obtain the theoretical shear stress:
\begin{align}
\sigma &\approx -P\int_{0}^{2\pi}d\theta \left[ b\sin^{2}(2\theta)-c\cos^{2}(2\theta) \right] \nonumber \\
&=-\pi P(b-c).
\label{eq:theo2}
\end{align}

\begin{figure}[htbp]
  \centering
    \includegraphics[clip,width=8.5cm]{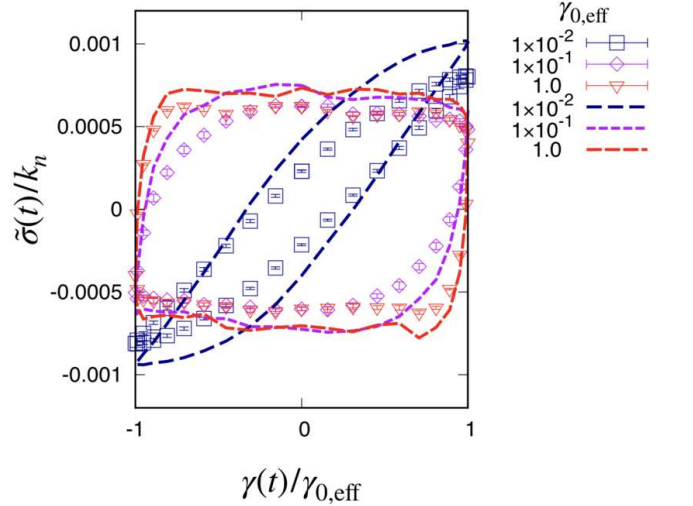}
    \caption{
    Plots of stress-strain (Lissajous) curves for various $\gamma_{0,\textrm{eff}}$ at $\mu=1.0$ and $\hat P=2.0\times10^{-3}$.
    Numerical data are plotted by open symbols.
    Symmetric stress by phenomenology $\sigma^{\textrm{sym}}(t)$ with aid of Eqs. \eqref{eq:theo} and \eqref{eq:theo2} are plotted with lines.
}
\label{fig:Lissajous}
\end{figure}

The stress-strain (Lissajous) curve for large strain amplitudes is plotted in Fig. \ref{fig:Lissajous},
where the stress from our simulation and the phenomenology from Eq. (\ref{eq:theo2}) are expressed by the symbols and the lines, respectively.  
It is remarkable that the phenomenological evaluation of the stress $\sigma$ from Eqs. \eqref{eq:theo} and \eqref{eq:theo2} recovers the stress-strain curve for $\gamma_{0,\textrm{eff}}/\hat{P} \ge 1$, at least.

Because we can reproduce the stress-strain curve  via phenomenology with the aid of $\zeta_{N}(\theta)$ and $\zeta_{T}(\theta)$,
we can also reproduce $G'$ and $G''$ via the phenomenological storage and loss moduli with the aid of Eqs.~\eqref{G'} and \eqref{G''}.
The lines in Figs. \ref{fig:sGs}(b) and \ref{fig:sGl}(b) show the phenomenological $G'$ and $G''$, respectively.
Although the phenomenology cannot be used for small strain amplitude, we reproduce the quantitative behavior of $G'$ and $G''$ for $\gamma_{0,\textrm{eff}}\gtrsim \hat P$.

\vspace*{0.4cm}
\section{Scaling laws for various friction coefficients\label{Apppsca}}

\begin{figure}[htbp]
  \centering
    \includegraphics[clip,width=8.5cm]{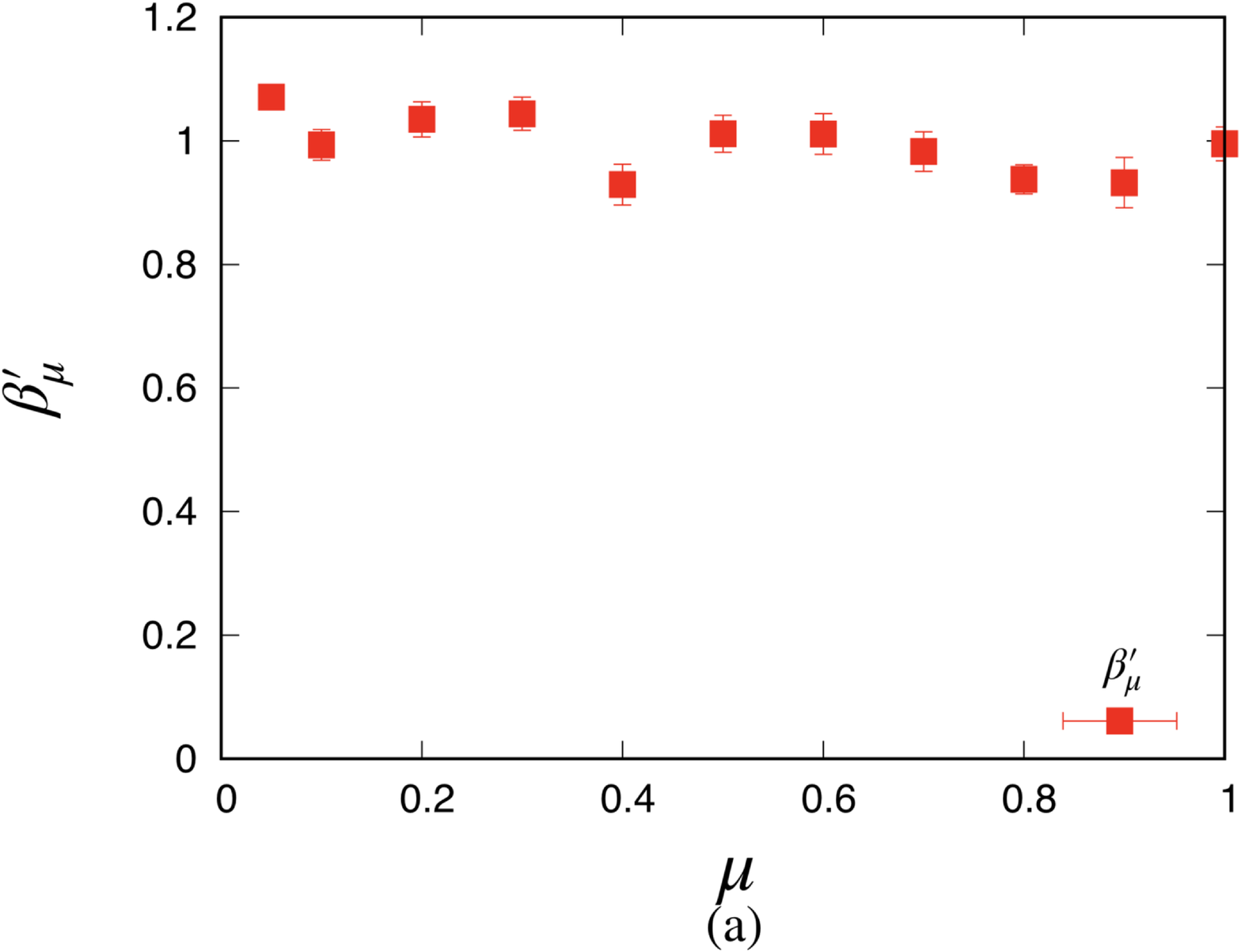}
    \includegraphics[clip,width=8.5cm]{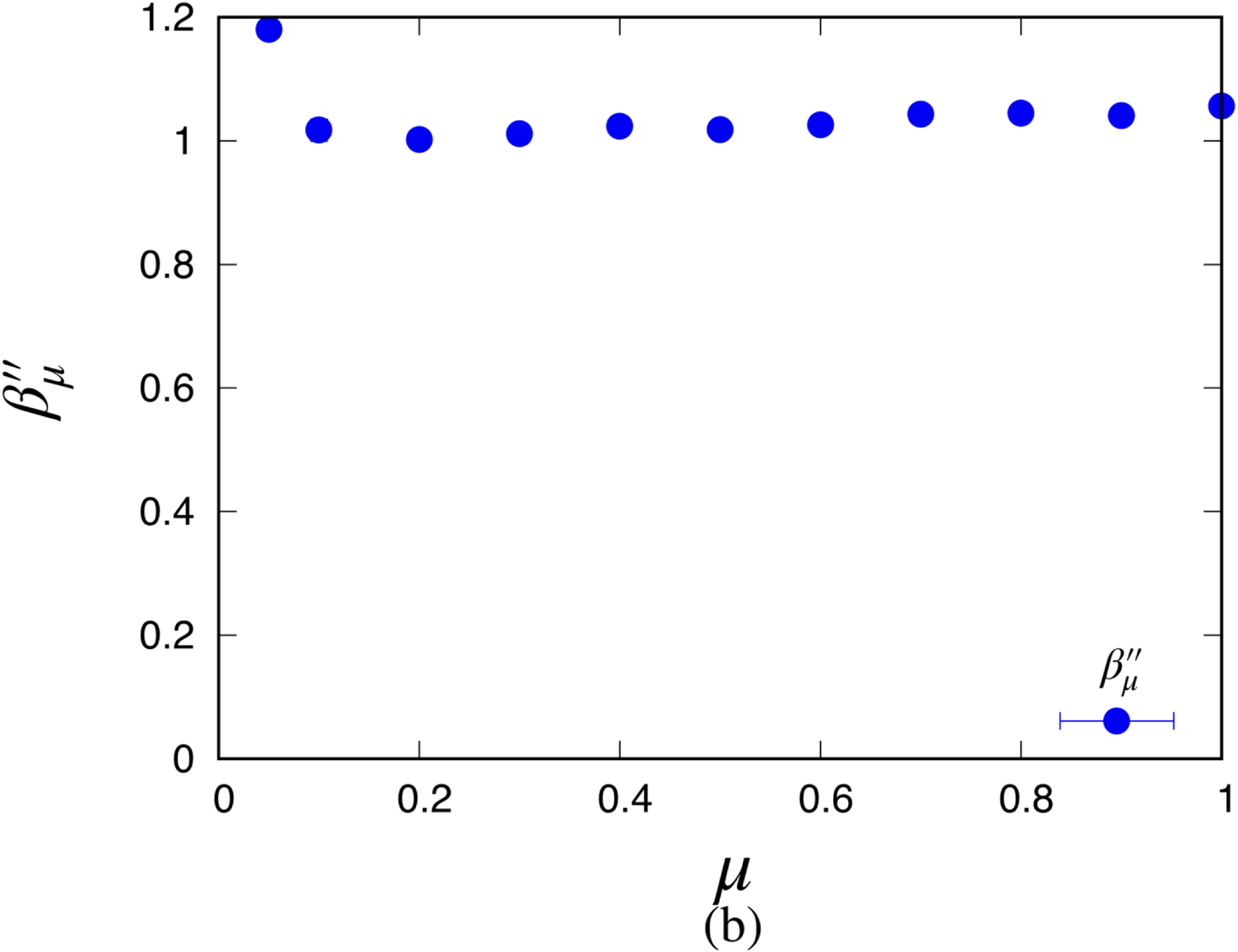}
    \caption{Plots of (a) $\beta_{\mu}'$ and (b) $\beta_{\mu}''$ against $\mu$. }
    \label{fig:expo}
\end{figure}

We have discussed the scaling laws only for $\mu=1.0$ in the previous section.
In this section we discuss how the results of Eqs. (\ref{eq:sgs1})-(\ref{eq:sgl3})  depend on $\mu$.
In short,
(i) the scaling laws in Eqs. (\ref{eq:sgs1})-(\ref{eq:sgl3}) still hold for arbitrary $\mu$, and
(ii) the exponents $\beta_{\mu}'$ and $\beta_{\mu}''$ are independent of $\mu$.

For arbitrary $\mu$ the storage modulus satisfies the relation:
\begin{align}
G'&=G'_{\textrm{res}}( \hat P)\mathcal{G'}\left( \frac{\gamma_{0,\textrm{eff}}}{\hat P^{\beta_{\mu}'}} \right),\\
G_{\textrm{res}}'(\hat P):&=\lim_{\gamma_{0,\textrm{eff}}\to 0}G'(\gamma_{0,\textrm{eff}}, \hat P), \\
\lim_{x\to 0}\mathcal{G'}(x)&=1,\ \lim_{x\rightarrow\infty}\mathcal{G'}(x)\sim x^{-1}.
\end{align}
and the loss modulus satisfies the relation:
\begin{align}
G''&=G'_{\textrm{res}}(\hat{P})\mathcal{G''}\left( \frac{\gamma_{0,\textrm{eff}}}{\hat P^{\beta''_{\mu}}} \right),\\
\lim_{x\to 0}\mathcal{G''}(x)&=\textrm{const.},\ \lim_{x\rightarrow\infty}\mathcal{G''}(x)\sim x^{-1} .
\end{align}
The estimated scaling exponents from our simulation are plotted in Fig. \ref{fig:expo}.
The exponents $\beta'_{\mu}$ and $\beta''_{\mu}$ are almost independent of $\mu$, which are approximately equal to unity.

\begin{figure}[htbp]
  \centering
    \includegraphics[clip,width=8.5cm]{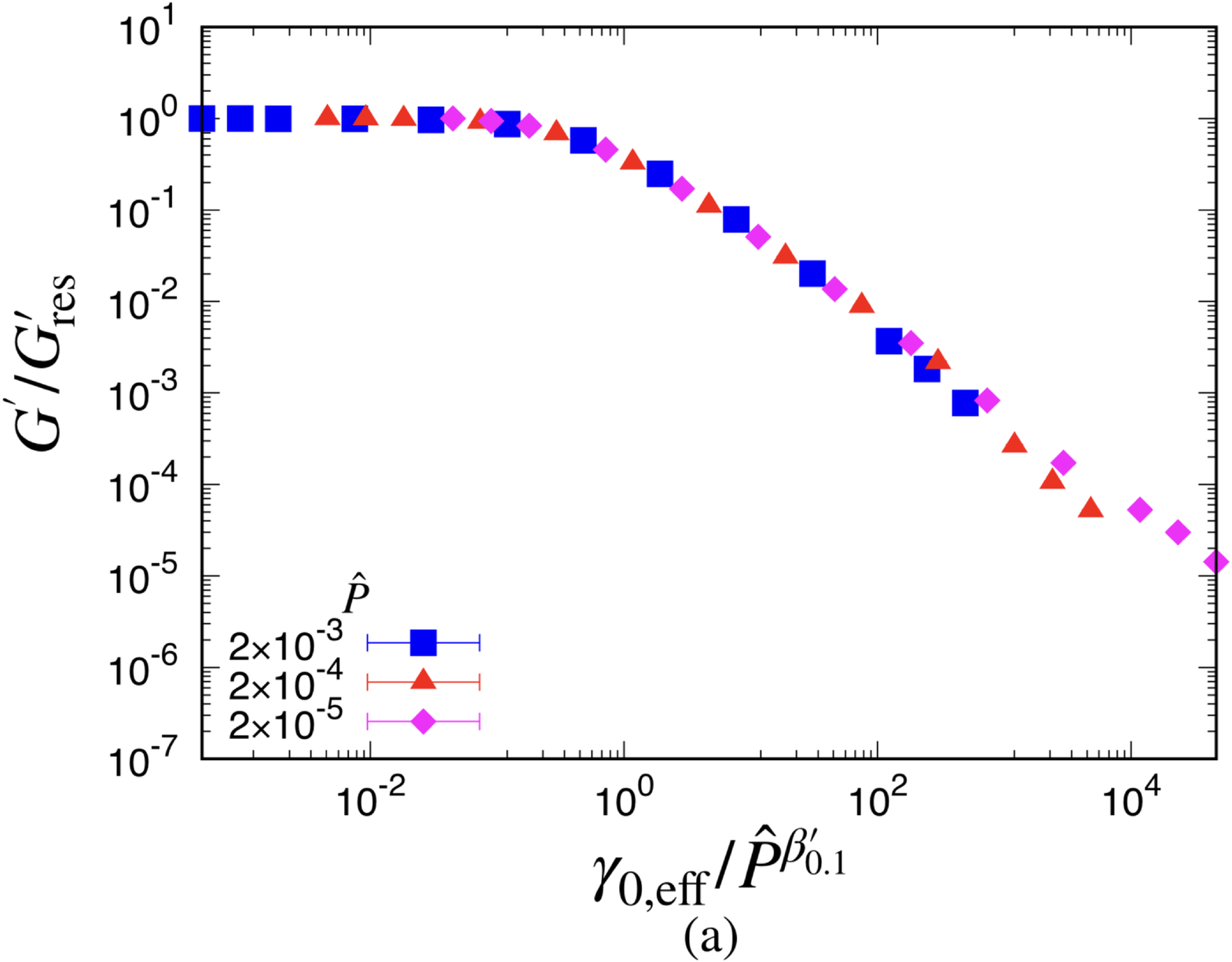}
    \includegraphics[clip,width=8.5cm]{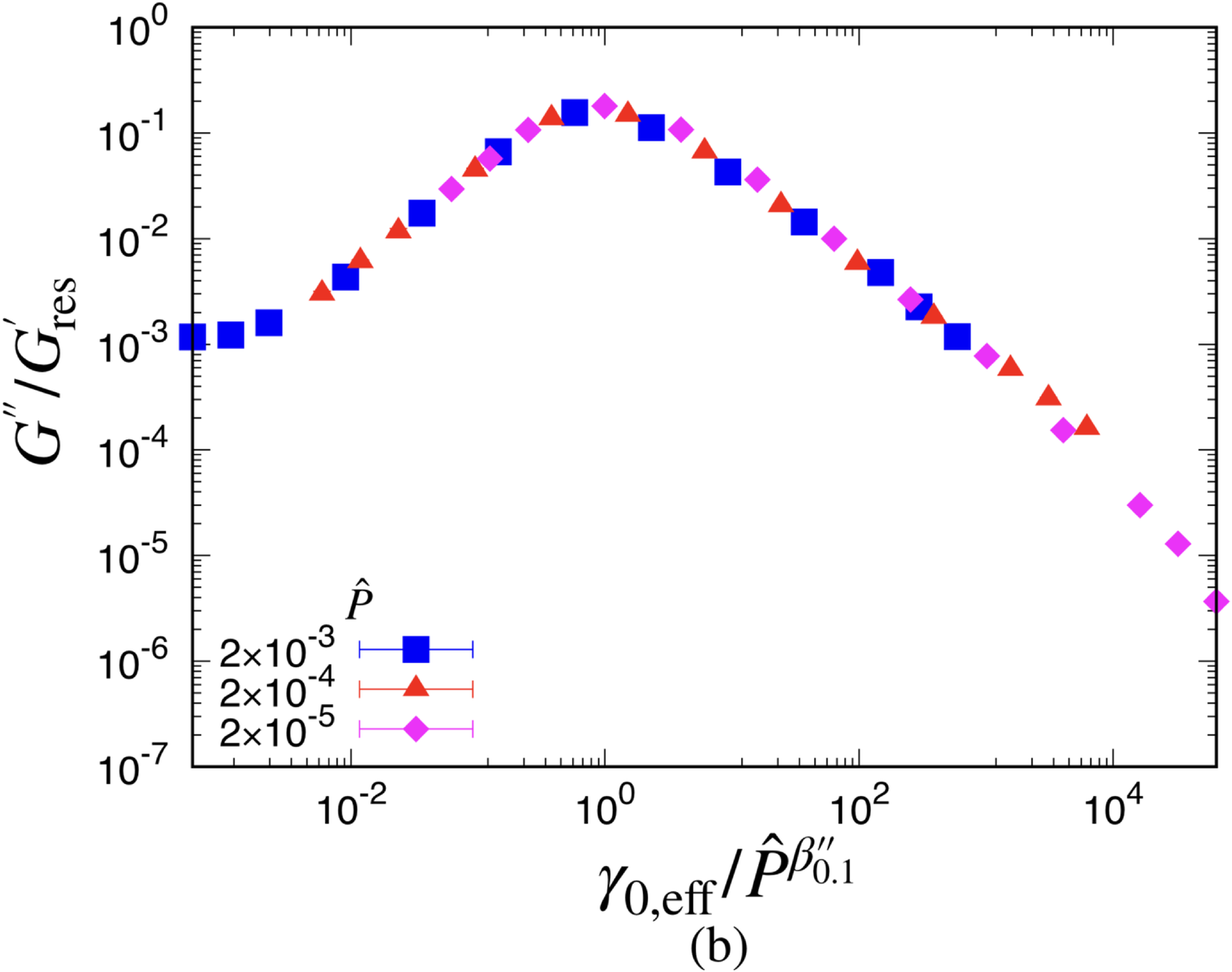}
    \caption{Scaling plots of (a) $G'$ and (b) $G''$ against scaled strain amplitude for various dimensionless pressures $\hat P$ at $\mu=0.1$}
    \label{fig:sgsls}
\end{figure}

The scaling plots of $G'$ and $G''$ against the scaled the strain amplitude at $\mu=0.1$ are presented in Fig. \ref{fig:sgsls}.
These scaling plots suggest that the scaling discussed in Sec.~\ref{sec3} can be used for arbitrary $\mu$.

\begin{figure}[htbp]
  \centering
    \includegraphics[clip,width=8.5cm]{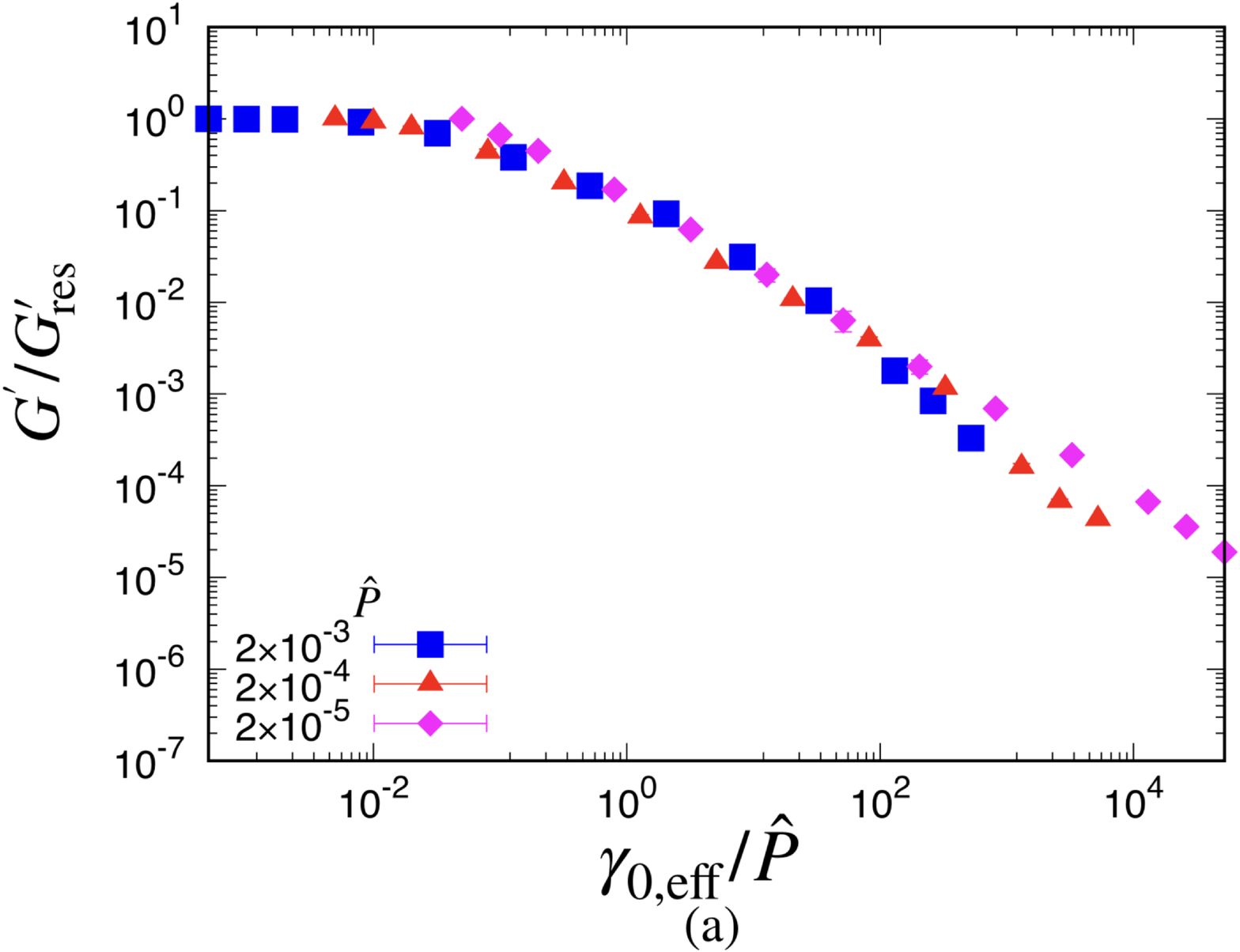}
    \includegraphics[clip,width=8.5cm]{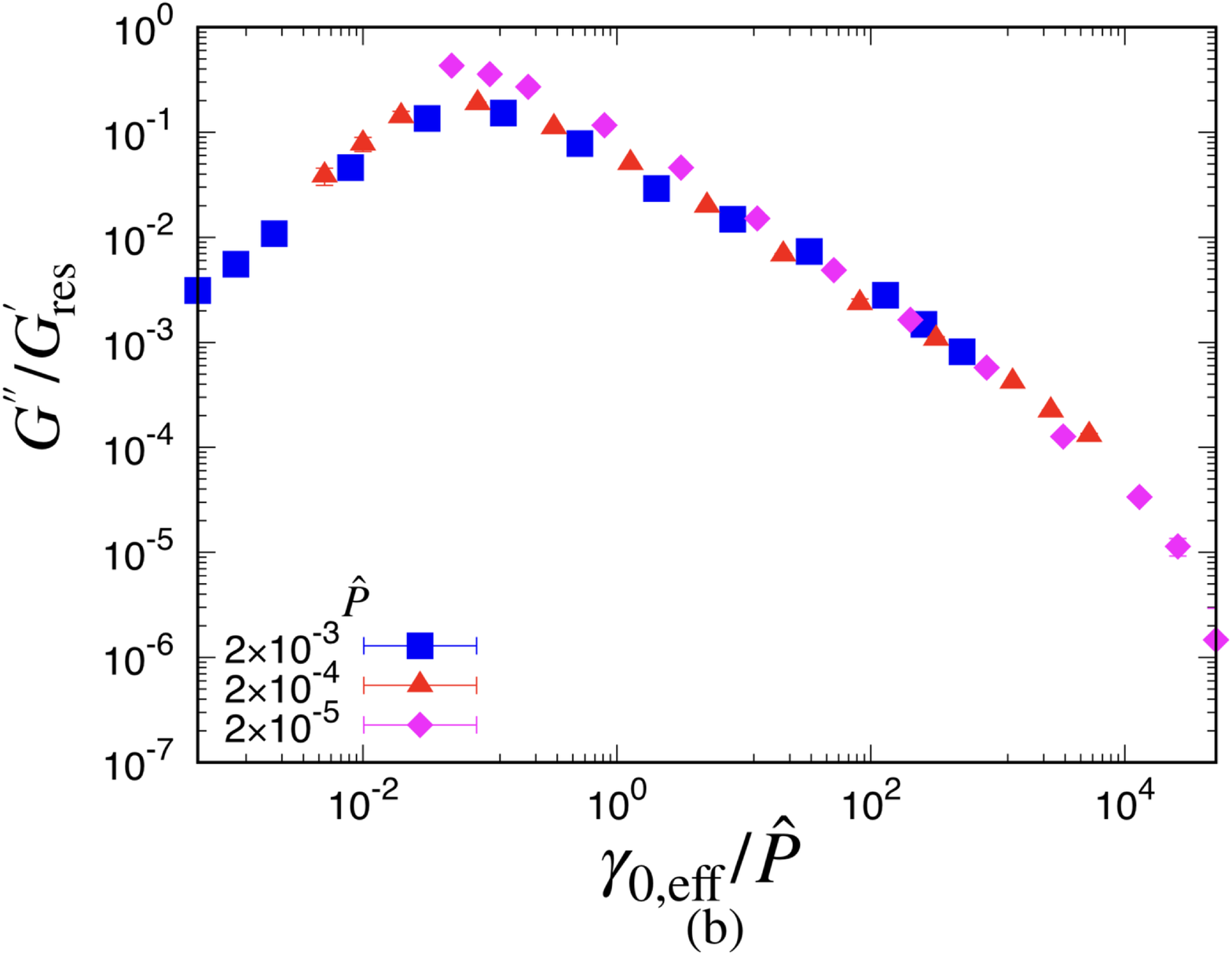}
    \caption{
    Scaling plots of (a) $G'$ and (b) $G''$ against scaled strain amplitude for various dimensionless pressures $\hat P$ at $\mu=0.01$.
    }
    \label{fig:sgslss}
\end{figure}

To confirm our conjecture that the scaling is independent of $\mu$ we examine the case of $\mu=0.01$.
It is interesting that the scaling laws seem to be valid even for $\mu=0.01$ without any changes of scaling exponents, though the distinction between the plastic and elastic regimes is not sharp in this case (see Fig. \ref{fig:sgslss}).
Nevertheless, we should emphasize that the results of the small friction coefficient limit for frictional grains is completely different from those in frictionless cases (see Appendix \ref{Appmu0})
in which the scaling laws do not exist as presented in Ref.~\cite{BSosc}.
This singularity in the zero-friction limit is consistent with that observed in Ref.~\cite{OHdst2}.

\section{discussion and conclusion\label{dandc}}

First, we can observe the dilatancy during the process because the volume can be changed.
It is known that the packing fraction $\delta\phi:=\phi-\phi_0$ from the unbiased fraction $\phi_0$ satisfies $\delta\phi\propto -\dot\gamma\sqrt{m/P}$ in a two-dimensional pressure control system under steady simple shear\cite{Cruz,Fdil}.
It is also known that the volume is compactified if the strain is small~\cite{Nowak98}.
The detailed behavior of the dilatancy will be reported elsewhere.

Second, the origin of the macroscopic friction law, i. e. $\sigma/P$ tends to be independent of the strain for large strain limit should be clarified.
This is equivalent to the microscopic determination of $b$ and $c$ in Eq. \eqref{eq:theo2}.
Quantitative analysis to determine $b$ and $c$ will be considered elsewhere.

In summary, we have numerically studied a frictional granular system confined by a constant pressure under oscillatory shear.
We confirmed the existence of scaling laws for the storage and loss moduli.
We also verified the absence of the distinction between the softening and yielding regimes for the frictional system.
These results are completely different from those for frictionless systems;
however, the friction coefficient can be disregarded, particularly for $0.01\le \mu \le 1$.
We found the crossover of the storage modulus from $G'/G'_{\textrm{res}}=1$ for small strain amplitude to $G'/G'_{\textrm{res}}\propto \hat P/\gamma_{0,\textrm{eff}}$ for large strain amplitude.
We also found that the scaling of the loss modulus which has three regimes, in which
(i) $G''$ is independent of the strain amplitudes for small strain amplitudes, 
(ii) $G''$ is a linear function of  $\gamma_{0,\textrm{eff}}$ for intermediate strain amplitudes,
and (iii) $G'' \sim \hat P/\gamma_{0,\textrm{eff}}$ for large strain amplitudes.
The phenomenological theory with the aid of the angular distributions of the contacts works in the plastic regime

\vspace*{0.4cm}
\noindent
{\bf Acknowledgement}

One of the authors (DI) expresses his sincere gratitude to T. G. Sano, S. Takada, and K. Saitoh for their assistance with the granular simulation.
The authors thank M. Otsuki, T. Kawasaki, R. Seto, and N. Oyama for fruitful discussions and usefull comments.
This work is partially supported by the Grant-in-Aid of MEXT for Scientific Research (Grant No. 16H04025) and the Programs YITP-T-18-03 and YITP-W-18-17.

\vspace*{0.4cm}

\appendix

\section{Coupled stress and asymmetric stress \label{Appasym}}

\begin{figure}[htbp]
  \centering
    \includegraphics[clip,width=7cm]{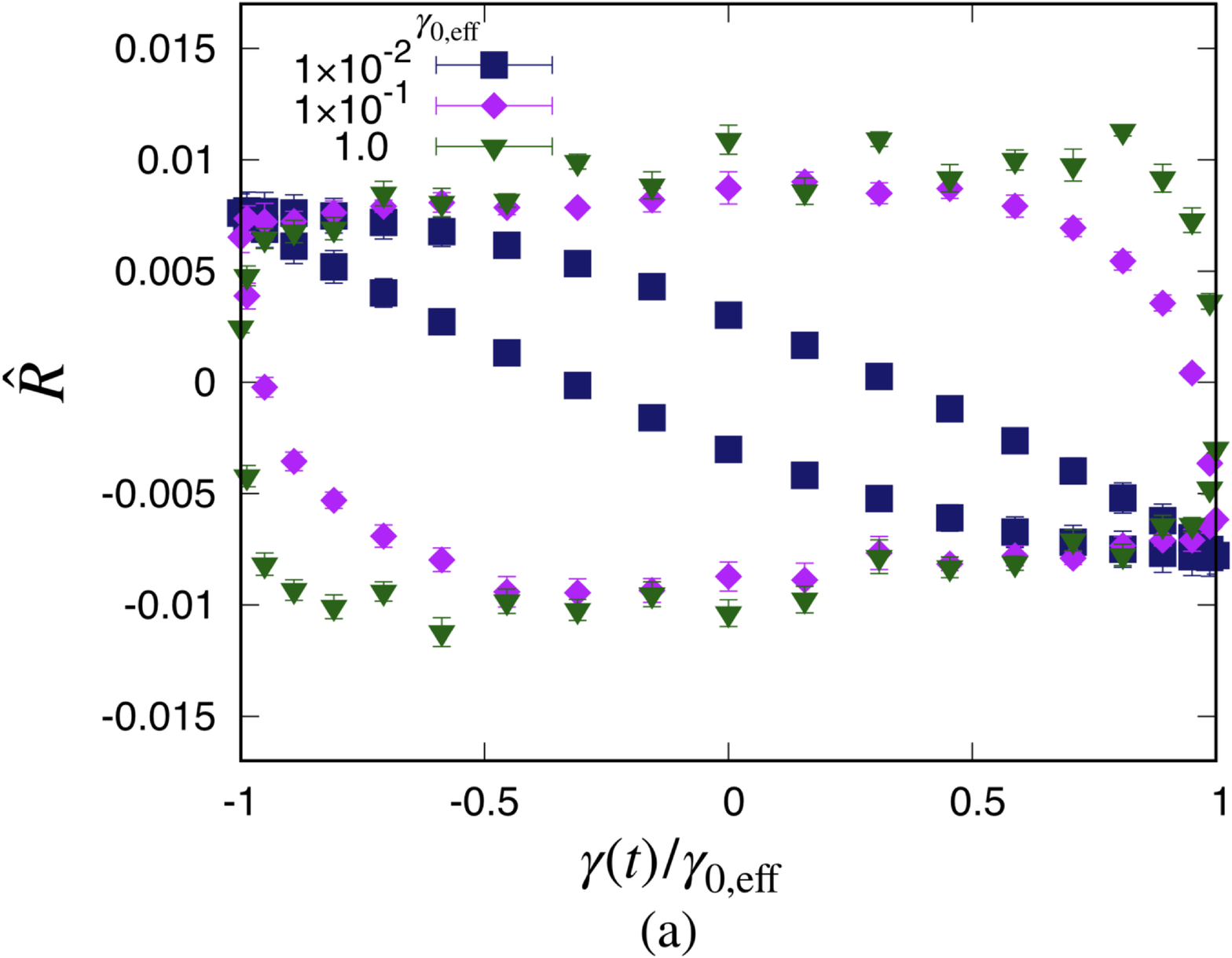}
    \includegraphics[clip,width=7cm]{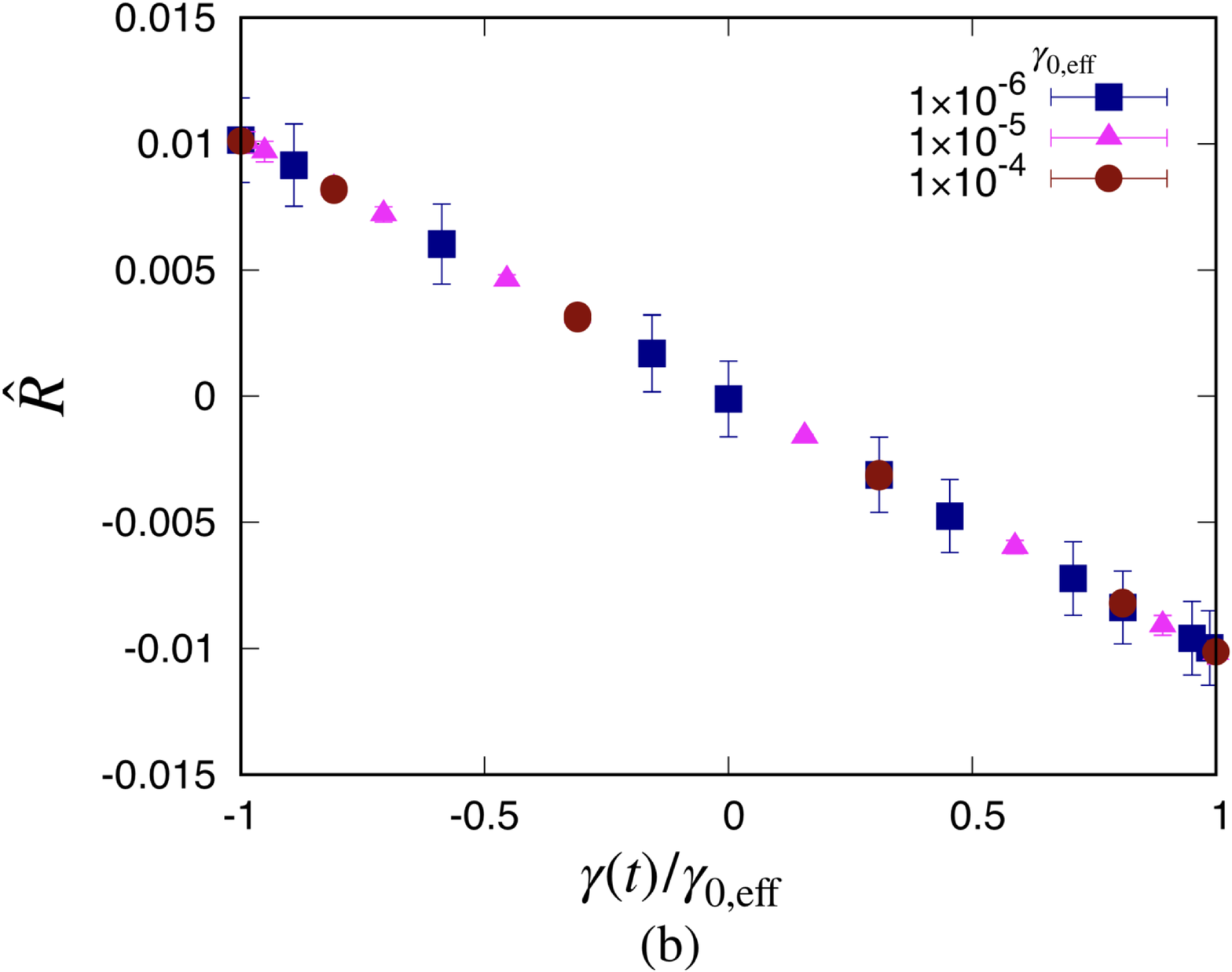}
    \caption{
    Plots of $\hat R$ against $\gamma(t)/\gamma_{0,\textrm{eff}}$ at $\mu=1.0$ and $\hat P=2.0\times 10^{-3}$  for (a) large strain amplitudes and (b)  small strain amplitudes.
    }
    \label{fig:asym}
\end{figure}

\begin{figure}[htbp]
  \centering
    \includegraphics[clip,width=7cm]{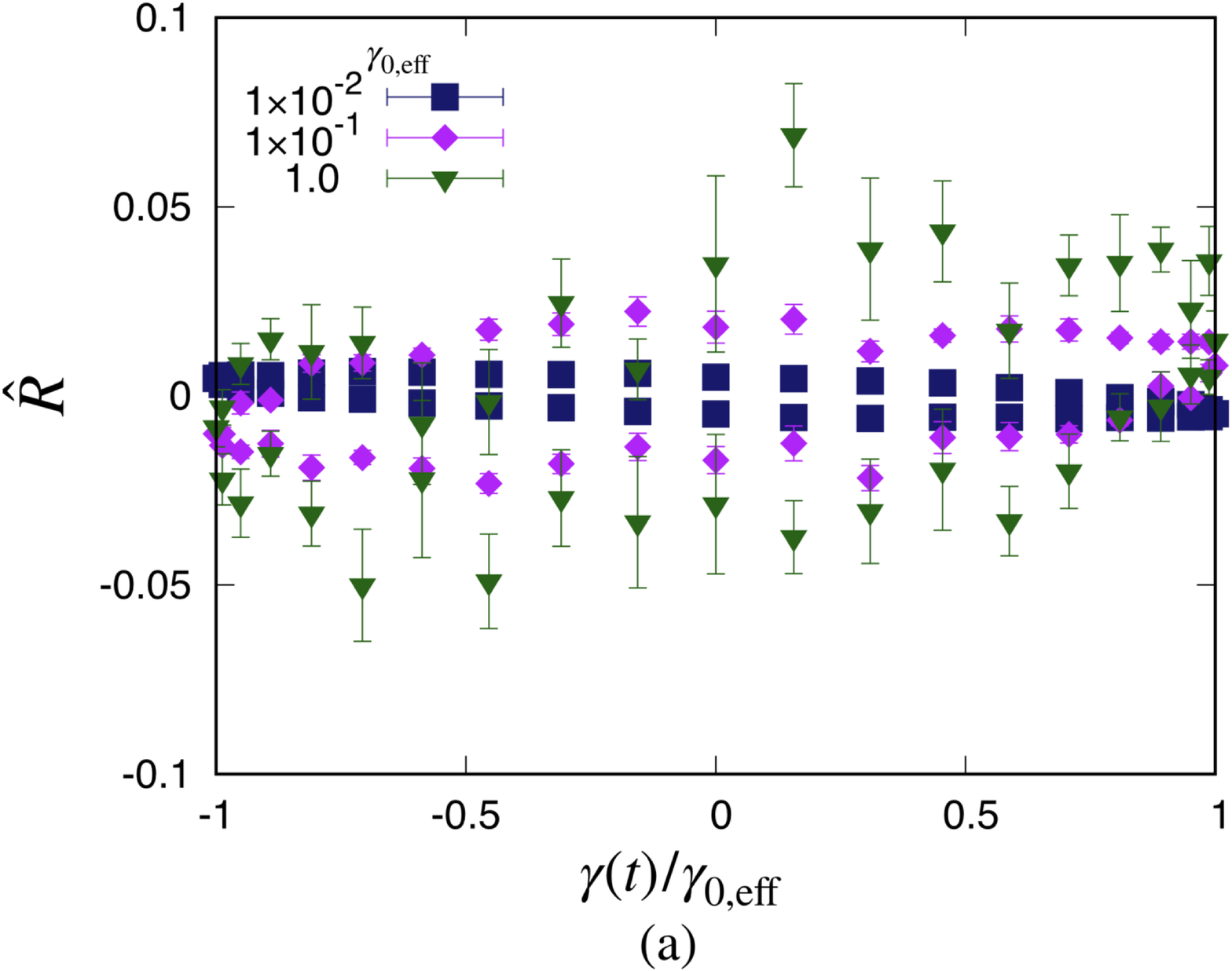}
    \includegraphics[clip,width=7cm]{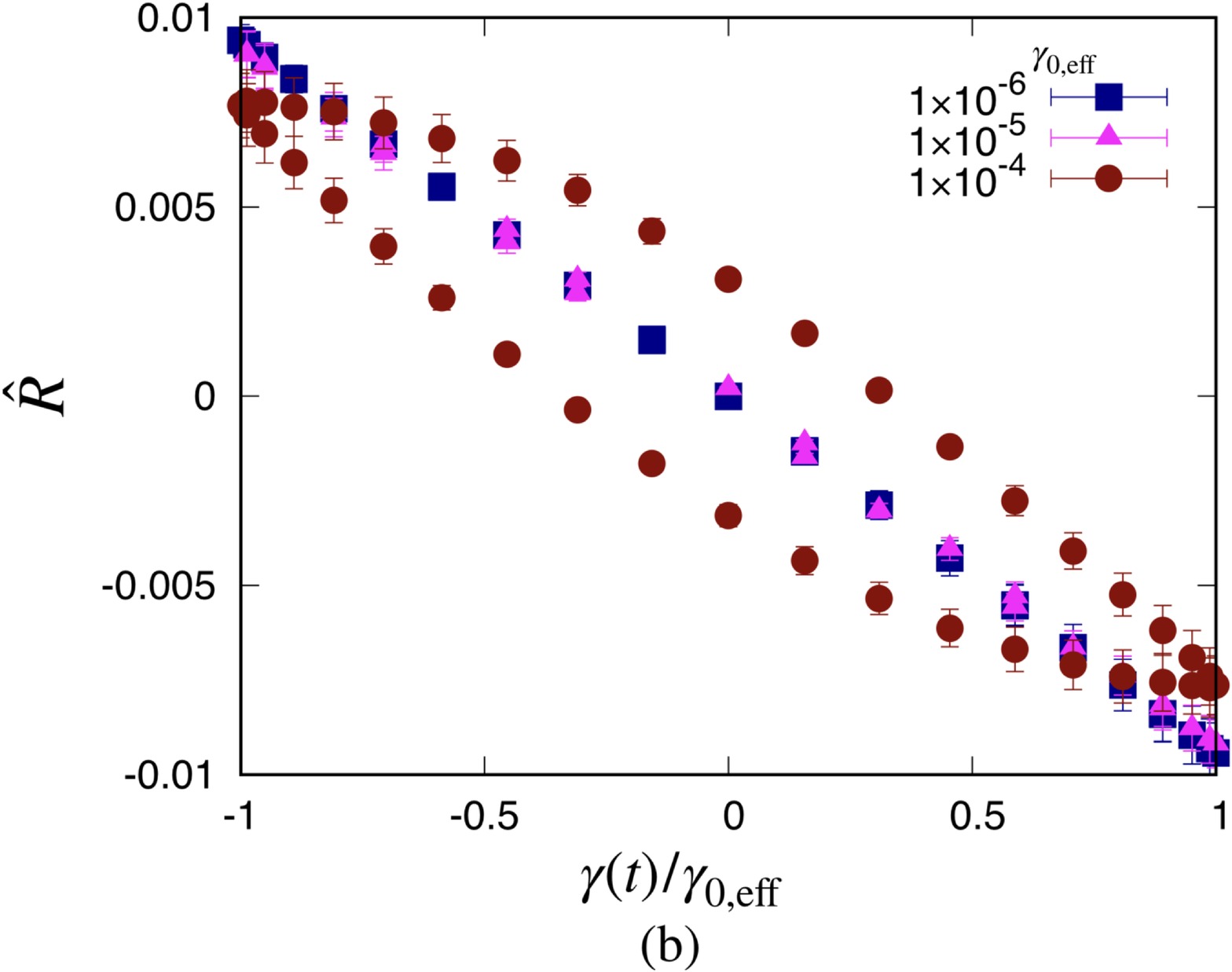}
    \caption{
    Plots of $\hat R$ against $\gamma(t)/\gamma_{0,\textrm{eff}}$ at $\mu=1.0$ and $\hat P=2.0\times 10^{-5}$  for (a) large strain amplitudes and (b) small strain amplitudes.
    }
    \label{fig:asym2}
\end{figure}

In this appendix, we discuss the coupled stress $R$ which is the asymmetric part of the stress tensor~\cite{Gol} defined as
\begin{align}
R:=&\frac{1}{L_{x} L_{y}}\sum_{i} \sum_{j>i}T_{ij},
\end{align}
where $T_{ij}:=x_{ij}f_{ij,y}-y_{ij}f_{ij,x}$.
Here, $f_{ij,k}$ with $k=x$ or $k=y$, $x_{ij}$, and $y_{ij}$ are the $k-$component of $\bm{f}_{ij}$, $x-$component of $\bm{r}_{ij}$, and $y-$component of $\bm{r}_{ij}$, respectively.  
Let us introduce a normalized coupled stress $\hat R$:
\begin{align}
\hat R:=\frac{\tilde R}{\tilde\sigma(\Omega t=\pi/4)},
\end{align}
where $\tilde R:=R-\int_{0}^{\tau_{p}}dt R/\tau_{p}$.
We have confirmed $\hat R \ll 1$ for high pressure as shown in Fig. \ref{fig:asym},
though $\hat{R}(t)$ exhibits a stress-strain curve similar to the symmetric stress in the main text.
Although $\hat{R}(t)$ for large strain amplitudes is visible, the largest value of $\hat R$ is 0.07 in Fig. \ref{fig:asym2} (a).
Therefore, we can safely ignore the contribution of the asymmetric stress tensor or the couple stress.

\section{Angular frequency dependence of $G'$ and $G''$~\label{Appome}}

\begin{figure}[htbp]
  \centering
    \includegraphics[clip,width=8.5cm]{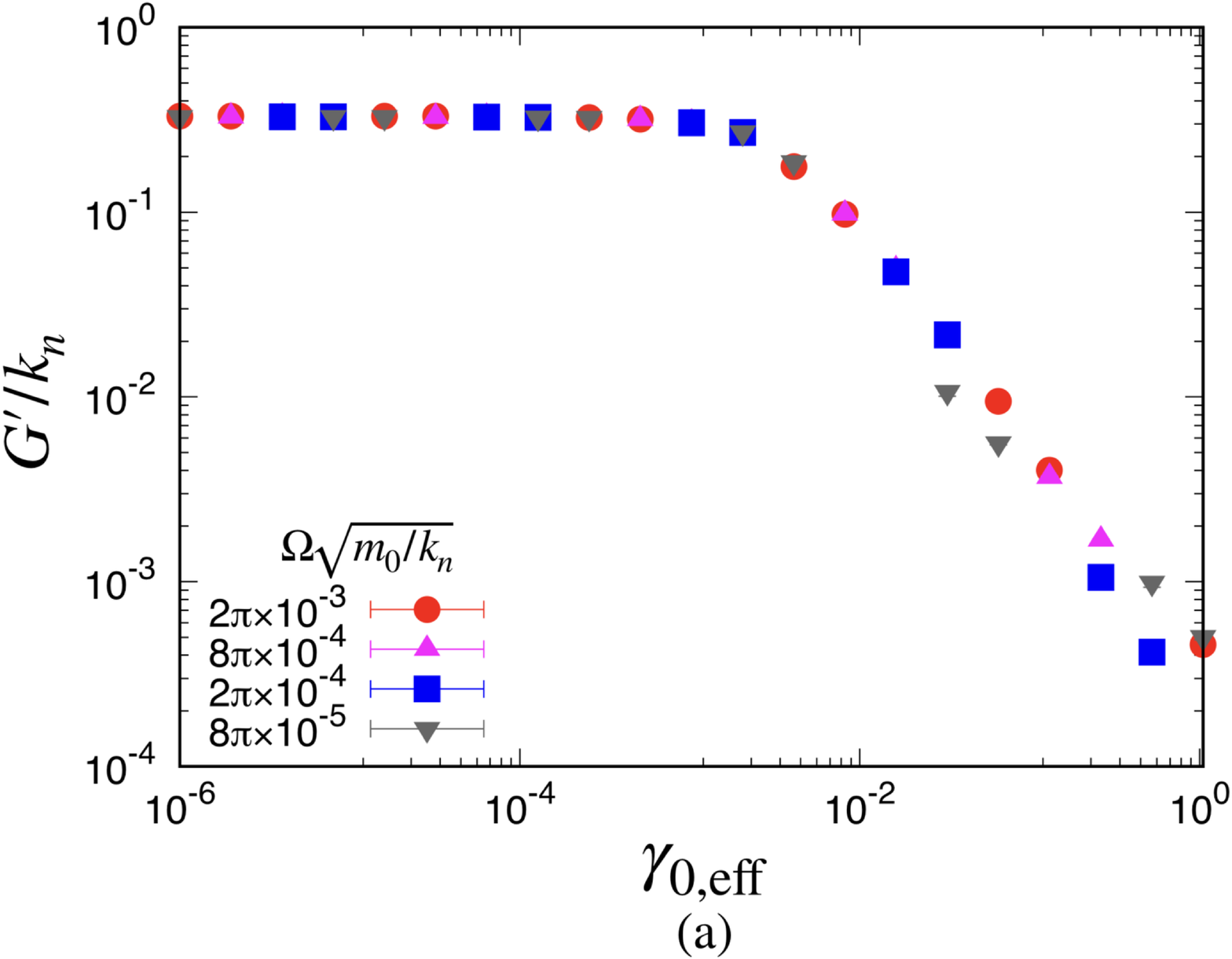}
    \includegraphics[clip,width=8.5cm]{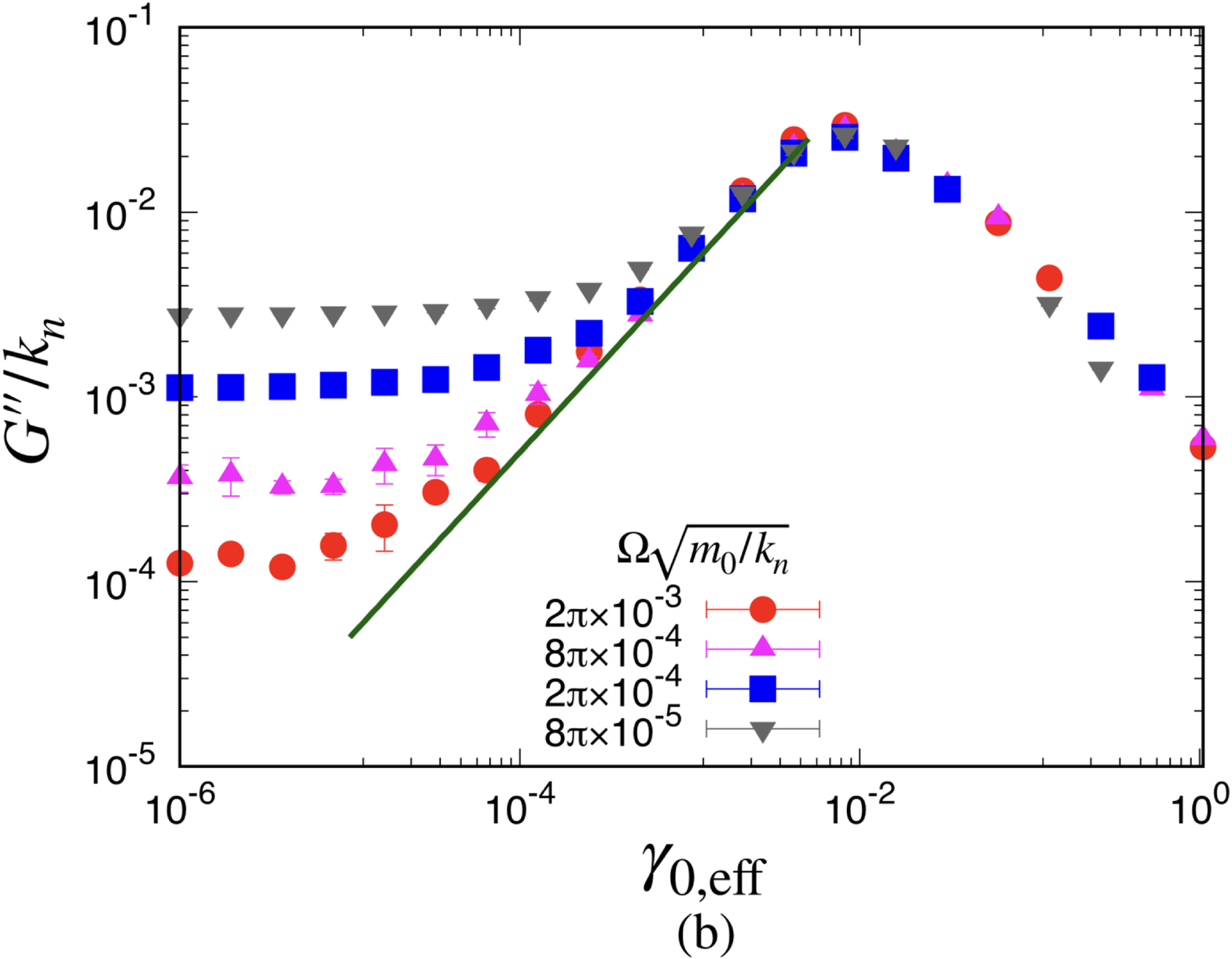}
    \caption{Plots of (a) $G'$ and (b) $G''$ against $\gamma_{0, {\textrm{eff}}}$ for various angular frequencies $\Omega$ at $\hat P=2.0\times10^{-3}$ and $\mu=1.0$. }
    \label{fig:Gslome}
\end{figure}

In this appendix, we investigate how the storage and loss moduli depend on $\Omega$ by fixing $\hat{P}=2.0 \times 10^{-3}$.
We have confirmed that $G'$ is almost independent of $\Omega$ for $\Omega\sqrt{m_0/k_n}/(2\pi) \le 10^{-3}$ in Fig. \ref{fig:Gslome} (a),
while $G''$ strongly depends on $\Omega$ for small $\gamma_{0,\textrm{eff}}/\hat{P}$ in Fig. \ref{fig:Gslome} (b).

\begin{figure}[htbp]
  \centering
    \includegraphics[clip,width=8.5cm]{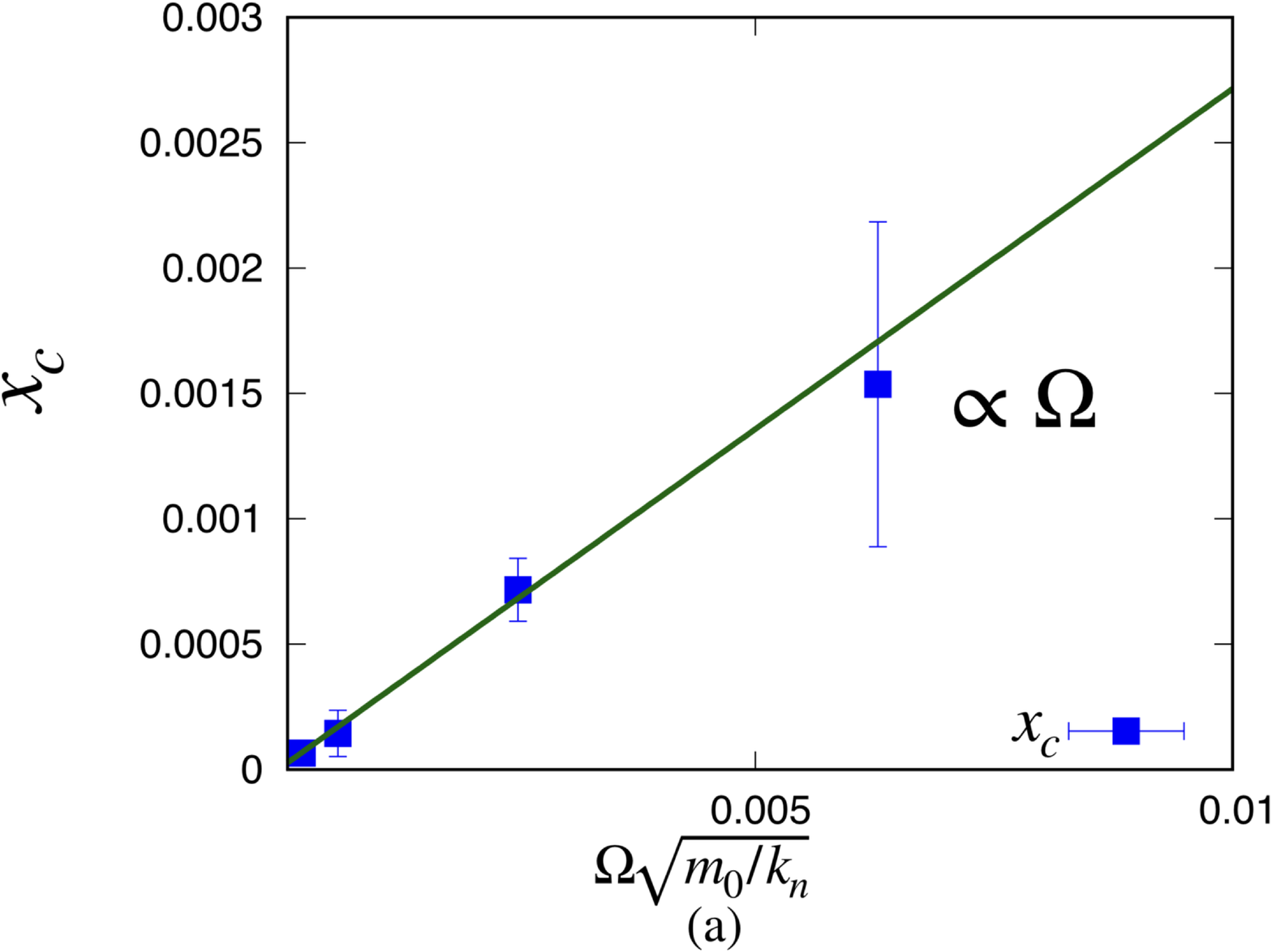}
    \includegraphics[clip,width=8.5cm]{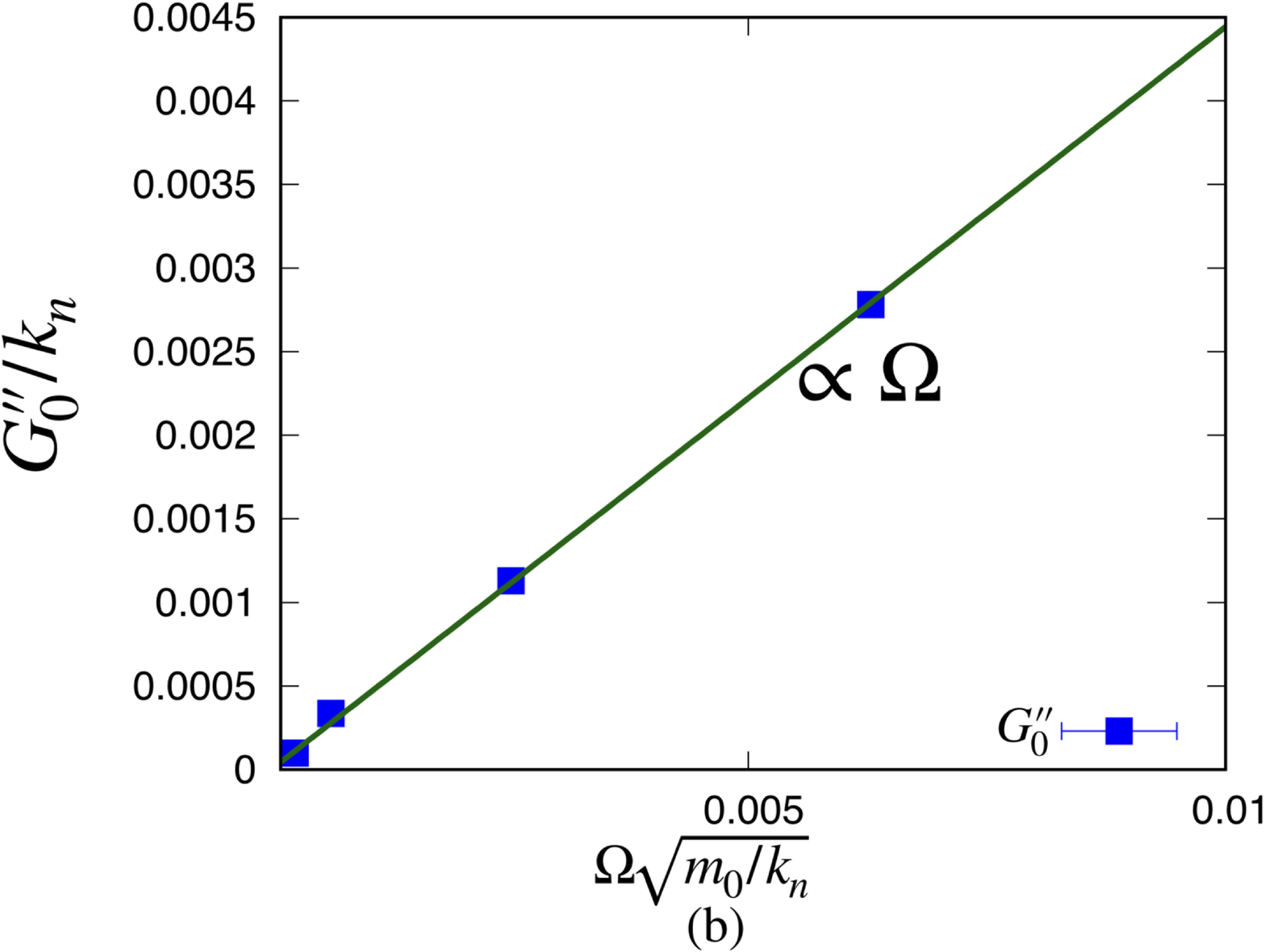}
    \caption{Plots of (a) crossover point $x_{c}$ for $G''$ against $\Omega$ and (b) $G''_0$ introduced in Eq.~\eqref{G_0} against $\Omega$.}
    \label{fig:xc}
\end{figure}

Here, we clarify how $G''$ depends on $\Omega$ for the small strain regime
($\gamma_{0,\rm{eff}}/\hat{P}<x_c$).
We have introduced a critical parameter $x_{c}$
as the middle point of $\gamma_{0,\textrm{eff}}/\hat{P}$ which satisfies
\begin{align}
0.4\le \frac{d\log G''}{d\log \gamma_{0,\textrm{eff}} }\le 0.8.
\end{align}
Figure~\ref{fig:xc} (a) shows that $x_c$ is also proportional to $\Omega$. 

We have also introduced $G''_{0}$ by
\begin{align}\label{G_0}
G''_{0}:=\lim_{\gamma_{0,\textrm{eff}}\to 0}G''.
\end{align}
From Fig. \ref{fig:xc}(b), we have confirmed that $G''_{0}\propto\Omega$, as mentioned in Sec.\ref{scaling_loss}.

\section{The loss modulus in the linear regime to $\gamma_{0,\rm{eff}}$\label{lossdim}}

In this appendix, we explain the origin of the regime (ii) in which $G''$ is proportional to $\gamma_{0,\rm{eff}}/\hat{P}$.
We also evaluate the value of $G''$ at the peak.

The behavior of $G''$ for $\gamma_{0,\textrm{eff}}/\hat P\lesssim 1$ can be interpreted as follows.
We will focus on the viscous shear stress $\sigma^{\textrm{(vis)}}$ for this purpose.
For simplicity, we ignore the difference between the peak point of $G''$ and the bending point of $G'$, 
although these two points are slightly different in reality.

Here the shear stress $\sigma^{\textrm{(vis)}}_{<}$ in the regime (ii) satisfying $\gamma_{0,\rm{eff}}\lesssim \hat{P}$ might obey
\begin{align}
\sigma^{\textrm{(vis)}}_{<}= f(\hat{P}) m_{0}\tau^{-2},
\end{align}
where $\tau$ is a typical time scale for elastic energy relaxation, and
the subscript $<$ represents the regime $\gamma_{0,\textrm{eff}}/\hat P\lesssim 1$.
We can assume that the function $f(\hat{P})$ obeys a power law as $f(\hat{P})\sim \hat{P}^\Delta$.

The estimation of the time scale $\tau$ is as follows.
The strain energy $E$ for a grain with the stiffness $k_n$ under the strain $\gamma$ might be expressed as $E\sim k_n (\gamma \ell)^2$, where $\ell$ stands for a typical length scale. This restoring energy should be balanced with the kinetic energy $E\sim m (\ell/\tau)^2$ where $m$ and $\tau$ are the mass of the grain and the typical time scale to characterize the time differentiation.
Therefore, we obtain $\tau\sim \sqrt{m_0/k_n}/\gamma$.
In our case, we can replace $\gamma$ by $\gamma_{0,\rm{eff}}$.
We then obtain the form
\begin{align}\label{sigma<}
\sigma^{\textrm{(vis)}}_{<}\sim k_n \hat{P}^\Delta \gamma_{0,\rm{eff}}^2 ,
\end{align}

For the plastic regime $\gamma_{0,\textrm{eff}}/\hat P\gtrsim 1$, on the other hand, the stress $\sigma^{\textrm{(vis)}}_{>}$ ($>$ stands for the regime
$\gamma_{0,\textrm{eff}}/\hat P \gtrsim 1$) satisfies
\begin{align}\label{sigma>}
\sigma^{\textrm{(vis)}}_{>}\sim P.
\end{align}
The stresses in Eqs.~\eqref{sigma<} and \eqref{sigma>} must take the identical value as
\begin{align}
\sigma^{\textrm{(vis)}}_{<}=\sigma^{\textrm{(vis)}}_{>}
\end{align}
at the peak of $G''$ under the condition $\gamma_{0,\rm{eff}}\sim \hat{P}$.
Therefore, we obtain the relation $\hat{P}^{\Delta+1}\sim 1$.
Because the right hand side of this equation is independent of $\hat{P}$, we get $\Delta=-1$,
and $\sigma^{\rm{(vis)}}
\sim k_n\gamma_{0,\rm{eff}}^{2}/\hat{P} \sim P
$ at the peak.
Finally, we obtain $G''_{\textrm{max}}=k_{n}$.
This indicates that the peak value of the loss modulus should be independent of $P$.

\section{Fitting parameters for $\zeta_{N}$ and $\zeta_{T}$}\label{fitting}

\begin{figure}[htbp]
  \centering
    \includegraphics[clip,width=8.5cm]{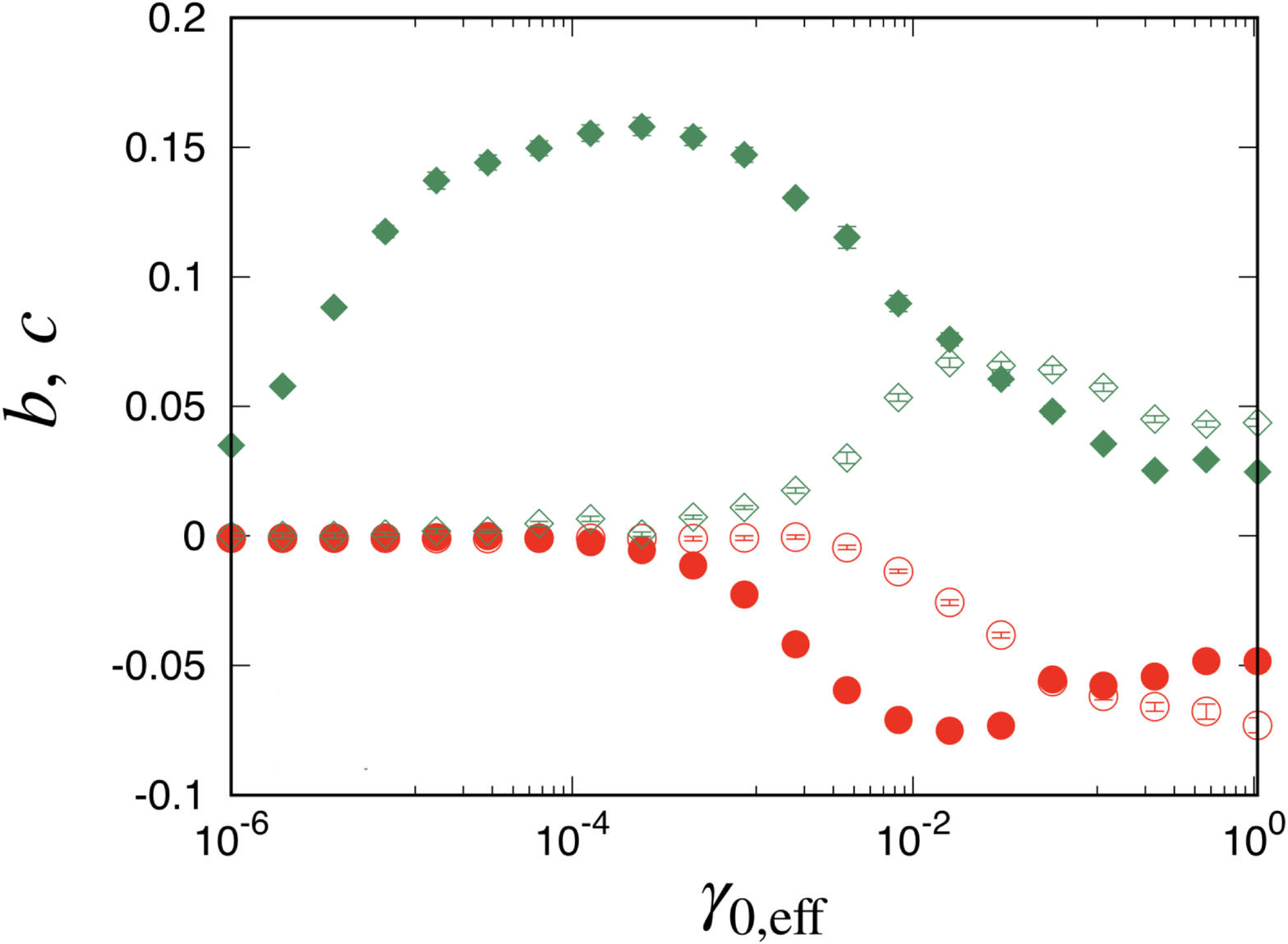}
\caption{
Plot of fitting parameters $b$ (circles) and $c$ (diamonds) as functions of 
$\gamma_{0,\textrm{eff}}$ at $\hat P=2.0\times10^{-3}$, where fitting data at $\Omega t=(2n+1/2)\pi$ with  $n=1,\ \cdots,\ 9$ are plotted as filled symbols and fitting data at $\Omega t=2n\pi$ with $n=1,\ \cdots,\ 9$ are plotted as open symbols.
}
    \label{fig:paraS}
\end{figure}

\begin{figure}[htbp]
  \centering
    \includegraphics[clip,width=8.5cm]{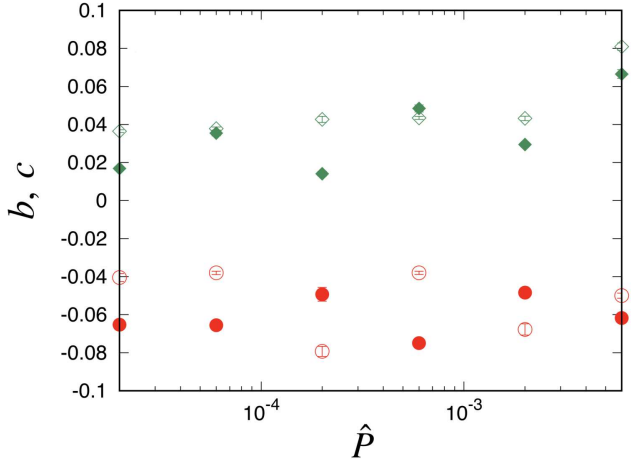}
\caption{Plot of fitting parameters $b$ (circles) and $c$ (diamonds) as functions of $\hat P$ at $\gamma_{0,\textrm{eff}}=1.0$, 
where fitting data at $\Omega t=(2n+1/2)\pi$ with $n=1,\ \cdots,\ 9$ are plotted as filled symbols and fitting data at $\Omega t=2n\pi$ with $n=1,\ \cdots,\ 9$ are plotted as open symbols.
}
    \label{fig:paraP}
\end{figure}

In this appendix, we present how the fitting parameters $b$ and $c$ introduced in Fig. \ref{fig:afd} and Eq. \eqref{eq:theo2} depend on $\hat P$ and $\gamma_{0,\textrm{eff}}$.
Here, we numerically evaluate the strain dependence of $b$ and $c$ at $\hat P=2.0\times10^{-3}$ in Fig. \ref{fig:paraS}.
We also plot the pressure dependence of $b$ and $c$ at $\gamma_{0,\textrm{eff}}=1.0$ in Fig. \ref{fig:paraP}.

\section{Results for frictionless grains}\label{Appmu0}

\begin{figure}[htbp]
  \centering

    \includegraphics[clip,width=8.5cm]{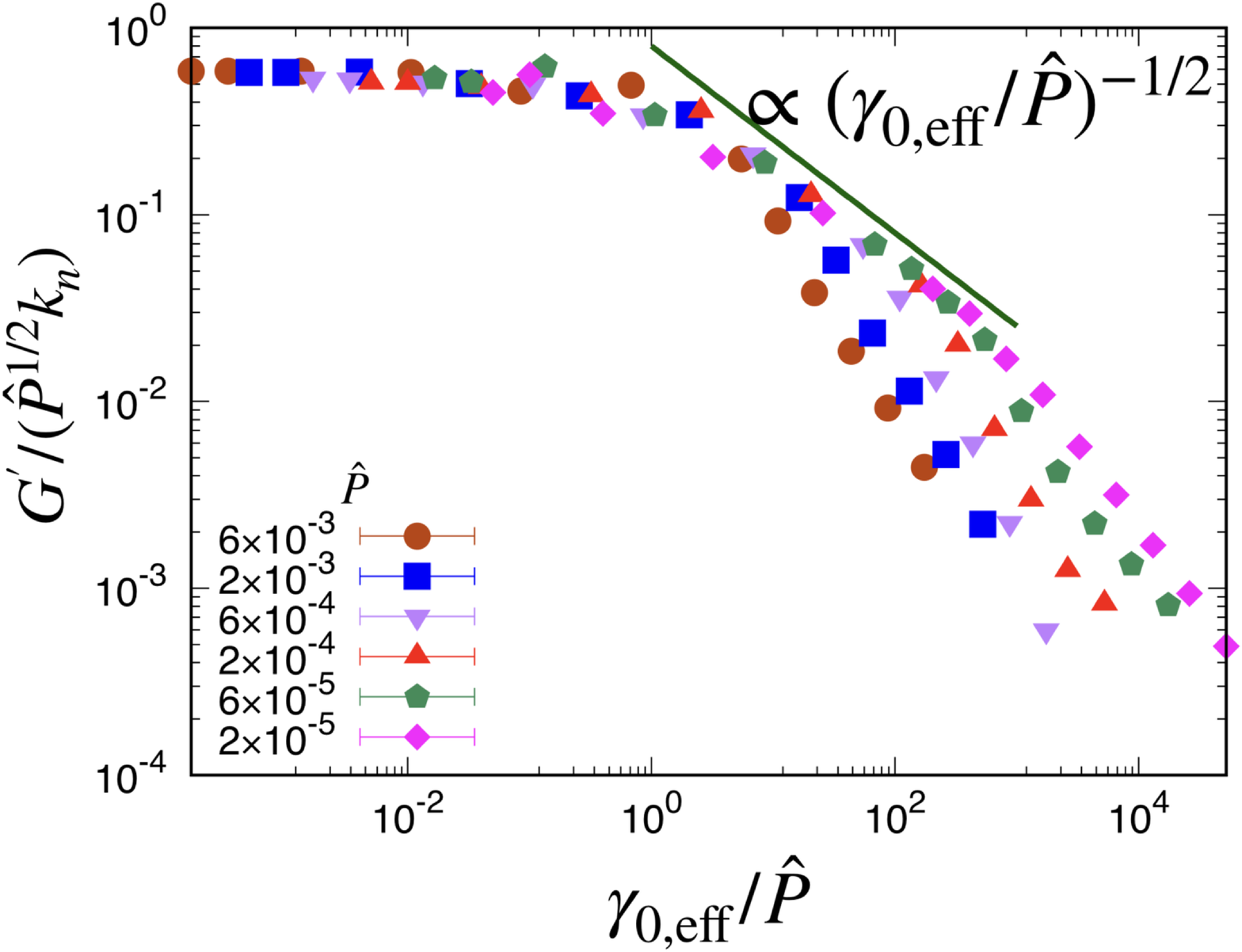}
    \caption{Plot of the scaled $G'$ against $\gamma_{0,\textrm{eff}}/\hat P$ for frictionless systems.}
    \label{fig:Gs0}
\end{figure}

In this appendix, we present the results of our simulation for the storage and loss moduli for a frictionless system.
It is known that the rigidity of two-dimensional frictionless systems near the jamming point is proportional to $\hat P^{1/2}$ \cite{Djam,Ojam,Otsuki14}.
Because of $P\propto \phi-\phi_J$ near the jamming point, 
$G'/(\phi-\phi_J)^{1/2}$ corresponds to $G'/\hat P^{1/2}$, 
where $\phi$ and $\phi_J$ are the volume fraction and the jamming fraction, respectively. 
We expect that the scaling between $G'/\hat P^{1/2}$ and $\gamma_{0. \rm {eff}}$ is held corresponding to Ref.~\cite{Otsuki14}. 
Here, we plot $G'/\hat P^{1/2}$ against $\gamma_{0,\textrm{eff}}/\hat P$ in Fig. \ref{fig:Gs0}.
However, there is no universal scaling law for frictionless systems;
this is significantly different from the results of finite $\mu$ (even for $\mu=0.01$) which exhibit definite scaling laws. 
The scaling corresponding to Ref.~\cite{Otsuki14} is only visible for a weakly plastic regime under low pressure.
Figure \ref{fig:Gs0} suggests the existence of the softening region when $\gamma_{0,\textrm{eff}}/\hat P\simeq 1$ for low pressure cases.
On the other hand, we find the yielding for $\gamma_{0,\textrm{eff}}/\hat{P}>10^3$.
Notably, the existence of two distinct regions corresponds to the softening and yielding described by Ref.~\cite{BSosc}, which contains the background friction.

\begin{figure}[htbp]
  \centering
   \includegraphics[clip,width=8.5cm]{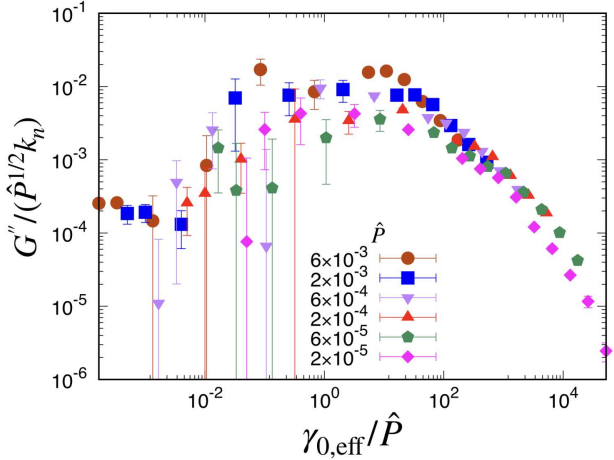}
   \caption{Scaling plot of $G''$ for frictionless systems.}
   \label{fig:Gl0}
\end{figure}

We plot $G''$ against $\gamma_{0,\textrm{eff}}/\hat P$ in Fig. \ref{fig:Gl0}.
The scaling for $G''$ is also absent, which is completely different from that of $\mu=0.01$.

\begin{thebibliography}{99}
\bibitem{Gra}H. M. Jaeger, S. R. Nagel, and R. P. Behringer, Rev. Mod. Phys. \textbf{68}, 1259 (1996).
\bibitem{Col}P. N. Pusey, in Liquids, Freezing and the Glass Transition, Part II, Les Houches Summer School Proceedings Vol. 51, edited by J.-P. Hansen, D. Levesque, and J. Zinn-Justin (Elsevier, Amsterdam, 1991), Chap. 10.
\bibitem{For}D. J. Durian and D. A. Weitz, in Foams, in Kirk-Othmer Encyclopedia of Chemical Technology, 4th ed., edited by J. I. Kroschwitz (Wiley, New York, 1994), Vol. 11, p. 783.
\bibitem{Zhang05}H. P. Zhang, and H. A. Makse, Phys. Rev. E \textbf{72}, 011301 (2005).
\bibitem{LNjam}A. J. Liu and S. R. Nagel, Nature \textbf{396}, 21 (1998).
\bibitem{Djam}D. J. Durian , Phys. Rev. Lett. \textbf{75}, 4780 (1995).
\bibitem{Mjam1}H. A. Makse, N. Gland, D. L. Johnson and L. M. Schwartz, Phys. Rev. Lett. \textbf{83}, 5070 (1999).
\bibitem{Tjam}V. Trappe, V. Prasad, L. Cipelletti, P. N. Segre and D. A. Weitz, Nature \textbf{411}, 772 (2001).
\bibitem{Ojam}C. S. O'Hern, L. E. Silbert, A. J. Liu and S. R. Nagel, Phys. Rev. E \textbf{68}, 011306 (2003).
\bibitem{Mjam3}H. A. Makse, N. Gland, D. L. Johnson and L. Schwartz, Phys. Rev. E \textbf{70}, 061302 (2004).
\bibitem{Mjam}T. S. Majmudar, M. Sperl, S. Luding, and R. P. Behringer, Phys. Rev. Lett \textbf{98}, 058001 (2007).

\bibitem{Esca1}W. G. Ellenbroek, E. Somfai, M. van Hecke and W. van Saarloos, Phys. Rev. Lett. \textbf{97}, 258001 (2006).
\bibitem{Osca}P. Olsson and S. Teitel, Phys. Rev. Lett. \textbf{99}, 178001 (2007).
\bibitem{Hsca}T. Hatano, J. Phys. Soc. Jpn. \textbf{77}, 123002 (2008).
\bibitem{Wsca2}M. Wyart, H. Liang, A. Kabla and L. Mahadevan, Phys. Rev. Lett. \textbf{101}, 215501 (2008).
\bibitem{Esca2}W. G. Ellenbroek, M. van Hecke and W. van Saarloos, Phys. Rev. E \textbf{80}, 061307 (2009).
\bibitem{OHsca}M. Otsuki and H. Hayakawa, Phys. Rev. E \textbf{80}, 011308 (2009).
\bibitem{SHsca}K. Shundyak, M. van Hecke and W. van Saarloos, Phys. Rev. E \textbf{75}, 010301(R) (2007).
\bibitem{Ssca2}E. Somfai, M. van Hecke, W. G. Ellenbroek, K. Shundyak and W. van Saarloos, Phys. Rev. E \textbf{75}, 020301(R) (2007).
\bibitem{dst1} E. Brown and H. M. Jaeger, Phys. Rev. Lett. \textbf{103}, 086001 (2009).
\bibitem{Sisca}L. E. Silbert, Soft Matt. \textbf{6}, 2918 (2010).
\bibitem{OHdst1}M. Otsuki and H. Hayakawa, Phys. Rev. E \textbf{83}, 051301 (2011).
\bibitem{SMdst}R. Seto, R. Mari, J. F. Morris, and M. M. Denn, Phys. Rev. Lett. \textbf{111}, 218301 (2013).
\bibitem{dst2}M. Wyart and M. E. Cates, Phys. Rev. Lett. \textbf{112}, 098302 (2014).
\bibitem{OHdst2}M. Otsuki and H. Hayakawa, Phys. Rev. E \textbf{95}, 062902 (2017).
\bibitem{Kdst}T. Kawasaki and L. Berthier, Phys. Rev. E \textbf{98}, 012609 (2018).
\bibitem{Zsj}D. Bi, J. Zhang, B. Chakraborty and R. Behringer, Nature \textbf{480}, 355 (2011).
\bibitem{Exsj1}J. Zhang, T. Majmudar and R. Behringer, Chaos \textbf{18},041107 (2008)
\bibitem{Exsj2}J. Zhang, T. S. Majmudar, A. Tordesillas and R. P. Behringer, Granul. Matt. \textbf{12}, 159 (2010).
\bibitem{Exsj3}D. Wang, J. Ren, J. A. Dijksman, H. Zheng and R. P. Behringer, Phys. Rev. Lett. \textbf{120}, 208004 (2018).
\bibitem{Nusj1} S. Sarkar, D. Bi, J. Zhang, R. P. Behringer and B. Chakraborty , Phys. Rev. Lett. \textbf{111}, 068301 (2013).
\bibitem{Nusj2}S. Sarkar, D. Bi, J. Zhang, J. Ren, R. P. Behringer and B. Chakraborty, Phys. Rev. E \textbf{93}, 042901 (2016).
\bibitem{OHsj}M. Otsuki and H. Hayakawa, arXiv:1810.03846 (to be published in Phys. Rev. E).

\bibitem{Pradipto} Pradipto and H. Hayakawa, Soft Matter \textbf{16}, 945 (2020).

\bibitem{Fall15} A. Fall, F. Bertrand, D. Hautemayou, C. Mezi\`ere, P. Moucheront, A. Lema\^itre and G. Ovarlez, Phys. Rev. Lett. \textbf{114}, 098301 (2015).
\bibitem{Peters16} I. R. Peters, S. Majumdar and H. M. Jaeger, Nature \textbf{532}, 214 (2016).

\bibitem{Kumar} N. Kumar and S. Luding, Granul. Matter \textbf{18}, 58 (2016).
\bibitem{Jin17} Y. Jin and H. Yoshino, Nat. Commun. \textbf{8}, 14935 (2017).
\bibitem{Urbani} P. Urbani and F. Zamponi, Phys. Rev. Lett. \textbf{118}, 038001
(2017).
\bibitem{Jin18} Y. Jin, P. Urbani, F. Zamponi, and H. Yoshino, Sci. Adv.
{\bf 4}, eaat6387 (2018).


\bibitem{LAOS1}M. Doi and F. Edwards, The Theory of Polymer Dynamics (Oxford Univ. Press, 1986).
\bibitem{LAOS2}R. H. Ewoldt, A. E. Hosoi and G. H. McKinley, J. Rheol. \textbf{52}, 1427 (2008).
\bibitem{HyunRev} K. Hyun, M. Wilhelm, C. O. Klein, K. S. Cho, J. G. Nam, K. H. Ahn, S. J. Lee, R. H. Ewoldt and G. H. McKinley, Prog. Polym. Sci. \textbf{36}, 1697 (2011).
\bibitem{LAOS3}R. I. Argatov, A. Iantchenko and V. Kochebitov,  Contin. Mech. Thermodyn. \textbf{29}, 1375 (2017).

\bibitem{Tosc}B. P. Tighe, Phys. Rev. Lett. \textbf{107}, 158303 (2011).
\bibitem{Otsuki14} M. Otsuki and H. Hayakawa, Phys. Rev. E {\bf 90}, 042202 (2014).
\bibitem{BBosc}J. Boschan, D. V\aa gberg E. Somfai and B. P. Tighe , Soft Matt. \textbf{12}, 5450 (2016).
\bibitem{BSosc}S. Dagois-Bohy, E. Somfai, B. P. Tighe and M. van Hecke, Soft Matt. \textbf{13}, 9036 (2017).
\bibitem{BTosc}K. Baumgarten and B. P. Tighe , Soft Matt. \textbf{13}, 8368 (2017).

\bibitem{Rdil}O. Reynolds, Philos. Mag. Ser. \textbf{20}, 469 (1885).
\bibitem{Tdil}P. A. Thompson and G. S. Grest, Phys. Rev. Lett. \textbf{67}, 1751 (1991).
\bibitem{Sdil}J. H. Snoeijer, T. J. H. Vlugt, M. van Hecke and W. van Saarloos, Phys. Rev. Lett. \textbf{92}, 054302 (2004).
\bibitem{MiDi04}GDR MiDi , Eur. Phys. J. E \textbf{14}, 341 (2004).
\bibitem{Fdil}Y. Forterre and O. Pouliquen, Annu. Rev. Fluid Mech. \textbf{40}, 1 (2008).
\bibitem{Bdil}F. Boyer, \'E. Guazzelli and O. Pouliquen, Phys. Rev. Lett. \textbf{107}, 188301 (2011).
\bibitem{Crys}S. Luding, Phys. Rev. E \textbf{63}, 042201 (2001).
\bibitem{dem1}P. A. Cundall and O. D. L. Strack, Geotechnique \textbf{29}, 47 (1979).
\bibitem{dem2}S. Luding, Granul. Matt. \textbf{10}, 235 (2008).

\bibitem{Cruz}F. da Cruz, S. Emam, M. Prochnow, J. N. Roux and F. Chevoir, Phys. Rev. E. \textbf{72}, 021309 (2005).
\bibitem{rotTen}J.-J. Moreau, in Friction, Arching, Contact Dynamics, edited by D. E. Wolf and P. Grassberger (World Scientific, London, (1997), pp. 233-247.
\bibitem{Mitarai02} N. Mitarai, H. Hayakawa, and H. Nakanishi, Phys. Rev. Lett. {\bf 88}, 174301 (2002).
\bibitem{Gol}I. Goldhirsch, Granul. Matt. \textbf{12}, 239 (2010).
\bibitem{Hrev}M. van Hecke, J. Phys.: Condes. Matt. \textbf{22}, 033101 (2010).
\bibitem{calten1}F. Calvetti, G. Combe and J. Lanier, Mech. Cohesive-Frict. Mater. \textbf{2}, 121 (1997).
\bibitem{Kanatani}K. Kanatani, Powder Technol. \textbf{28}, 167 (1981).
\bibitem{calten3}B. Cambou, P. Dubujet, F. Emeriault and F. Sidoroff, Eur. J. Mech. A/Solids \textbf{14}, 255 (1995).
\bibitem{Nowak98}E. R. Nowak, J. B. Knight, E. Ben-Naim, H. M. Jaeger and S. R. Nagel, Phys. Rev E \textbf{57}, 1971 (1998).









\end{thebibliography}
\end{document}